\documentclass[aps,twocolumn,floats,prd,nofootinbib,superscriptaddress,10pt]{revtex4-1}

\usepackage[dvips]{graphicx} %
\usepackage{graphicx,amsmath,amsfonts,amssymb,slashed,float,hyperref}
\usepackage{bbold,wasysym}
\usepackage{graphicx}
\usepackage{array,multirow}
\usepackage[utf8]{inputenc}

\usepackage[usenames,dvipsnames]{xcolor} 

\usepackage{soul}

\definecolor{RedWine}{rgb}{0.743,0,0}
\definecolor{RoyalBlue}{rgb}{0.25,.41,.88}
\definecolor{celestialblue}{rgb}{0.29, 0.59, 0.82}

\setstcolor{Blue}

\definecolor{darkolivegreen}{rgb}{0.2, 0.5, 0.3}

\newcommand{\LCDM}{$\Lambda$CDM}
\newcommand{\PlanckTTTEEE}{{\sc Planck$^{\rm TTTEEE}$}}
\newcommand{\PlanckTTTEEELensing}{{\sc Planck$^{{\rm TTTEEE}+\phi\phi}$}}
\newcommand{\PlanckLensing}{{\sc Planck$^{\rm \phi\phi}$}}
\newcommand{\Planck}{{\sc Planck}}
\newcommand{\BAO}{{\sc BAO}}
\newcommand{\FS}{{\sc FS}}
\newcommand{\DES}{{\sc DES}}
\newcommand{\KV}{{\sc KiDS-VIKING}}
\newcommand{\Pantheon}{{\sc Pantheon}}
\newcommand{\SHOES}{{\sc SH0ES}}
\newcommand{\fEDE}{$f_{\rm EDE}(z_c)$} 
\newcommand{\Logzc}{${\rm Log}_{10}(z_c)$}

\usepackage[utf8]{inputenc}

\usepackage{amsthm,graphicx,multirow}
\usepackage{epsfig}
\usepackage{latexsym, amssymb} 
\usepackage{amsmath}
\usepackage{epstopdf}
\usepackage{hyperref}

\begin{document}

\title{The early dark energy resolution to the Hubble tension in light of weak lensing surveys and lensing anomalies}
\author{Riccardo Murgia}
\author{Guillermo F. Abell\'an}
\author{Vivian Poulin}

\affiliation{Laboratoire Univers \& Particules de Montpellier (LUPM), \\
            CNRS \& Universit\'e de Montpellier (UMR-5299), 
            Place Eug\`ene Bataillon, F-34095 Montpellier Cedex 05, France}
\date{January 2020}

\begin{abstract}
A constant Early Dark Energy (EDE) component contributing a fraction $f_{\rm EDE}(z_c)\sim 10 \%$ of the energy density of the universe around $z_c\simeq 3500$ and diluting as or faster than radiation afterwards, can provide a simple resolution to the Hubble tension,~the $\sim 5\sigma$ discrepancy -- in the $\Lambda$CDM context -- between the $H_0$ value derived from early- and late-universe observations. However, it has been pointed out that including Large-Scale Structure (LSS) data, which are in $\sim3\sigma$ tension with $\Lambda$CDM and EDE cosmologies, might break some parameter degeneracy and alter these conclusions.
We reassess the viability of the EDE against a host of high- and low-redshift measurements, by combining LSS observations from recent weak lensing (WL) surveys with CMB, Baryon Acoustic Oscillation (BAO), growth function (FS) and Supernova Ia (SNIa) data.
Introducing a model whose only parameter is \fEDE, we report in agreement with past work a $\sim 2\sigma$ preference for non-zero \fEDE~from \Planck~CMB data alone, while the tension with the local $H_0$ measurement from SH0ES is reduced below $2\sigma$. Adding BAO, FS and SNIa does not affect this conclusion, 
while the inclusion of a prior on $H_0$ from \SHOES~increase the preference for EDE over $\Lambda$CDM to the $\sim3.6\sigma$ level.
After checking the EDE non-linear matter power spectrum as predicted by standard semi-analytical algorithms via a dedicated set of $N$-body simulations, we test the 1-parameter EDE cosmology against WL data.  We find that it does not significantly worsen the fit to the $S_8$ measurement as compared to $\Lambda$CDM, and that current WL observations do not exclude the EDE resolution to the Hubble tension. We also caution against the interpretation of constraints obtained from combining statistically inconsistent data sets within the $\Lambda$CDM cosmology. In light of the CMB lensing anomalies, we show that the lensing-marginalized CMB data also favor non-zero \fEDE~at $\sim2\sigma$, predicts $H_0$ in $1.4\sigma$ agreement with \SHOES~and $S_8$ in $1.5\sigma$ and $ 0.8\sigma$ agreement with \KV~and \DES~respectively. There still exists however a $\sim2.5\sigma$ tension with the joint results from \KV~and~\DES. With an eye on Occam's razor, we finally discuss promising extensions of the EDE cosmology that could allow to fully restore cosmological concordance.
\end{abstract}
\date{\today}

\maketitle

\section{Introduction}

In recent years, several tensions between probes of the early- and late-universe have emerged. First and foremost, there exists a strong mismatch between the prediction of the current expansion rate of the universe (known as Hubble constant) in the $\Lambda$ cold dark matter ($\Lambda$CDM) model calibrated onto Planck CMB data, and its direct measurement using low redshift data (i.e., the classical distance ladder) \citep{Verde:2019ivm,2019NatRP...2...10R}. Originally, this ``Hubble tension'' was limited to the determination of the Hubble constant using type Ia supernovae by the SH0ES collaboration, whose latest {\em determination} is $H_0 = 74.03 \pm 1.42$ km/s/Mpc \citep{Riess:2019cxk}, while the {\em prediction} from the $\Lambda$CDM model inferred from Planck CMB data is $H_0 = 67.4\pm 0.5$ km/s/Mpc \citep{Aghanim:2018eyx}. In the last few years, tremendous progress have been made in measuring $H_0$ with alternative methods, such that nowadays there exist five other methods\footnote{These include strong-lens time delays of quasars \citep{Wong:2019kwg}, Tip of the red giant branch from the `CCHP' \citep{Freedman:2019jwv,Freedman:2020dne} (and re-evaluation by the \SHOES~team \citep{Yang:2018euj}), SNIa calibrated on Miras (an alternative to Cepheids) \citep{2020ApJ...889....5H}, water	masers	(sources	of	microwave	stimulated	emission)	
in	 four	 galaxies	 at	 great	 distances \citep{2020ApJ...891L...1P} and 	 Surface	 Brightness	
Fluctuations of distant galaxies \citep{Verde:2019ivm}.} to measure $H_0$ with few percent accuracy. Remarkably, various averages over these measurements (excluding correlated data) leads to $H_0$ values that ranges from $72.8\pm1.1$ and $74.3\pm1.0$, in $4.5$ to $6.3\sigma$ discrepancy with the prediction from \LCDM~ \cite{Verde:2019ivm,2019NatRP...2...10R}.
Similarly, it has been shown that the prediction from \LCDM~calibrated on any `early-universe' data (e.g.~\BAO+SNIa+BBN~\cite{Abbott:2017smn,Aylor:2018drw,Schoneberg:2019wmt,Arendse:2019hev}, WMAP+SPT and/or ACT \cite{Aylor:2018drw,Aiola:2020azj}) is always in good agreement with that of Planck.
A number of possible systematic effects affecting some of these measurements have been discussed~(see~e.g.~\cite{Rigault:2014kaa,Rigault:2018ffm,Blum:2020mgu,Millon:2019slk,Birrer:2020tax}), yet the existence of several vastly different methods -- none of which giving a value of $H_0$ smaller than $\sim70$ km/s/Mpc -- have triggered a wide range of theoretical activities to resolve the Hubble tension (see in particular \cite{Knox:2019rjx} for a recent review). Indeed, this tension between different measurements of the Hubble constant  could point to a major failure of the $\Lambda$CDM scenario, and hence to a new cosmological paradigm: that would be a new and unexpected breakthrough in cosmology. 
 
 
CMB data do not provide an absolute measurement of $H_0$. Rather, the value of $H_0$ is inferred within a given cosmological model  from a measurement of 
the angular scale of sound horizon $\theta_s\equiv r_s(z_*)/d_A(z_*)$, where $r_s(z_*)$ is the sound horizon at recombination  and  $d_A(z_*)$ is the angular diameter distance to recombination. The great challenge lies in that $\theta_s$ is nowadays measured at sub-percent-level accuracy with the latest CMB data~\cite{Aghanim:2018eyx}. This suggests two main ways of resolving the Hubble tension through new physics -- based on the requirement to keep the key angular scale $\theta_s$ fixed -- usually called {\em late-} and {\em early-}universe solutions. The first way boils down to changing the redshift evolution of the angular diameter distance in the late-universe, i.e. $z<z_*$, so as to force a higher $H_0$, {\em without} changing $d_A(z_*)$ nor $r_s(z_*)$. The second way amounts in reducing $r_s(z_*)$ in the early-universe, which automatically requires to reduce $d_A(z_*)$ by the same amount to keep $\theta_s$ fixed, that is most naturally done by increasing the value of $H_0$. A final, more subtle, way of resolving the $H_0$ tension comes from the fact that the position of the peaks receives an additional phase-shift from various effects, in particular from the gravitational pulling of CMB photons out of the potential wells by free-streaming neutrinos~\cite{2004PhRvD..69h3002B,Follin:2015hya,Baumann:2015rya}. Suppressing this phase-shift can change the value of $\theta_s$ deduced from a CMB power spectra analysis and in turn significantly increase $H_0$. 

 There have been many attempts to find extensions of the standard cosmological model, $\Lambda$CDM, which bring these estimates into agreement. However, theoretical explanations for the Hubble tension are not easy to come by.  Late-time observables, especially BAO and luminosity-distance to SNIa, place severe limitations on late-time resolutions \citep{Beutler:2011hx,Ross:2014qpa,Riess:2016jrr,DiValentino:2017zyq,Addison:2017fdm,DiValentino:2017iww,Alam:2016hwk,Bernal:2016gxb,Zhao:2017cud,Poulin:2018zxs,Aylor:2018drw,2020PhRvD.101h3524R,Alestas:2020mvb}. On the other hand, early-time resolutions affect the physics that determines the fluctuations in the CMB. At first sight, given the precision measurements of the CMB from Planck, this might appear to be even more constraining than the late-time probes of the expansion rate. Excitingly, there are a few early-time resolutions which do not spoil the fit to current CMB temperature measurements \cite{Poulin:2018cxd,Kreisch:2019yzn,Niedermann:2019olb,2020arXiv200401470E,2020arXiv200409487J}, sometimes even {\em improving it} over $\Lambda$CDM.

 EDE representing $\sim10\%$ of the total energy density of the universe around matter-radiation equality and diluting faster than radiation afterwards has been shown to provide a very good resolution to this tension. However, taken at face value, this model triggers a number of questions. On the theoretical side, it suffers from a strong coincidence problem as the fluid needs to become dynamical around a key era of the universe. This is not without reminding the standard coincidence problem of DE that such models were originally introduced to resolve. However, this coincidence might be the sign of a very specific dynamics to be uncovered; in fact there exist models in which the field becomes dynamical precisely around matter-radiation equality, either because of a phase-transition triggered by some other process (e.g.~the neutrino mass becoming of the order of the neutrino bath temperature \cite{Sakstein:2019fmf}) or because of a non-minimal coupling to the Ricci curvature \cite{Braglia:2020iik}.
 There exist also more concrete issues with Large-Scale Structure (LSS) observables, and in particular weak lensing (WL) surveys, which this 
 article aims to address. Indeed, a number of cosmic shear surveys (CFHTLenS \cite{Heymans:2013fya}, KiDS/Viking \cite{Hildebrandt:2018yau}, DES \cite{Abbott_2018}, HSC \cite{Hikage_2019}) have provided accurate measurements of the cosmological parameter $S_8\equiv \sigma_8(\Omega_m/0.3)^{0.5}$ -- where $\sigma_8$ measures the amplitude of fluctuations in a sphere of radius~$8$~Mpc$/h$ -- which are systematically lower than the $\Lambda$CDM prediction. The significance of this ``$S_8$ tension'' oscillates between 2 and 4$\sigma$ depending on the experiments, such that the discrepancy cannot easily be attributed to a statistical fluke. The problem for EDE cosmologies lies in the fact that the prediction for the value of $S_8$ is somewhat higher than that of $\Lambda$CDM. Therefore, taken at face values, these experiments pose a challenge to EDE cosmologies, and could exclude these models as a resolution to the Hubble tension \cite{Hill:2020osr}. A similar conclusion was reached in Refs~\cite{Ivanov:2020ril,DAmico:2020ods} with the inclusion of BOSS data in the effective field theory (EFT) of LSS framework.
 
 In this paper, we analyze the EDE cosmology resolving the Hubble tension in light of the latest \Planck~data (and the more precise polarization measurement) and confront it to the \KV~measurement of the cosmic shear power spectrum \cite{Asgari:2019fkq} and the joint measurement of $S_8$ from \KV+\DES\footnote{The re-analysis of BOSS data in the EFT of LSS is performed elsewhere~\cite{Smith:2020rxx}.}. 
 The \KV+\DES~measurements however rely on modelling the non-linear matter power spectrum on relatively small scales. This is done within numerical Einstein-Boltzmann solvers such as {\sc CAMB} \cite{Lewis:1999bs} or {\sc CLASS} \cite{Lesgourgues:2011re,Blas:2011rf}, through the {\sc halofit} \cite{Smith:2002dz,Takahashi:2012em} or {\sc HMcode} \cite{Mead:2015yca} algorithms, which have not been calibrated for EDE cosmologies. We thus check the predictions of these algorithms against the results of a set of dedicated cosmological $N$-body simulations, confirming that the qualitative departures from $\Lambda$CDM arising in the EDE cosmology are small enough to make use of these standard algorithms.
 We perform a series of Monte Carlo Markov Chain (MCMC) analyses with various combination of the latest Planck, BAO, growth factor and SNIa Luminosity distance measurements, the SH0ES measurement of $H_0$, and KiDS/Viking/DES data, in order to assess whether current observations exclude the EDE resolution to the Hubble tension.
 
 We find that, while the $S_8$ prediction from the best fit EDE cosmology is indeed $\sim 2.5\sigma$ higher than the measurement, KiDS data currently provide very little constraining power on the EDE parameters.
 Yet, it has been found in Ref.~\cite{Joudaki:2019pmv,Asgari:2019fkq} that a combination of KiDS and DES-Y1 (after re-calibration of the \DES~photo-metric redshifts) and  yields $S_8=0.755^{+0.019}_{-0.021}$, a result that is in 3.2$\sigma$ tension with Planck $\Lambda$CDM prediction\footnote{The joint analysis of KIDS1000+BOSS+2dfLenS has determined $S_8 = 0.766^{+0.020}_{-0.014}$ \cite{Heymans:2020gsg} in $3\sigma$ tension with Planck. Making use of these data would not affect our conclusions.}. 
 At such a level of discrepancy, one should be cautious when interpreting results obtained from combining Planck and WL data, even within $\Lambda$CDM. Indeed, we show that despite the inclusion of a Gaussian $S_8$ likelihood, the resulting cosmological model yields a very bad fit to the $S_8$ data, while providing very strong constraints on any parameter correlated with $S_8$ (e.g.~$\omega_{\rm cdm}$, $A_s$, $f_{\rm EDE}(a_c)$). It is easily conceivable that the resolution to the $S_8$ tension lies elsewhere (whether new physics related -- or not -- to the EDE, or systematic effects), such that any constraints derived from these combined data are artificial.
 
 It is also possible that this anomaly comes from CMB measurements. Interestingly, it has been found that there exists a lensing anomaly in Planck data (strengthened\footnote{See also Ref.~\cite{Efstathiou:2019mdh} for an independent re-analysis.} in the latest data released despite extensive efforts from the Planck team to pin down a possible systematic effect) \cite{Aghanim:2016sns,Aghanim:2018eyx}, which could be related to the $S_8$ tension \cite{Motloch:2018pjy,Motloch:2019gux}. Indeed,  the amplitude of the CMB lensing power spectrum deduced from the smoothing of the acoustic peaks at high-$\ell$'s is higher than that predicted by the $\Lambda$CDM cosmology obtained from the `unlensed' part of the CMB power spectrum\footnote{We will later on dub this the `unlensed' $\Lambda$CDM cosmology for simplicity.}. Moreover, it is also higher than that deduced from the lensing reconstruction such that, while this anomaly looks like lensing, it cannot be attributed to some extra source of CMB lensing. 
 Once marginalizing over the lensing information (which by itself is an interesting consistency check of the $\Lambda$CDM cosmology), it has been shown that the reconstructed cosmology has a smaller $A_s$ and $\omega_{\rm cdm}$ (as well as a higher $H_0$) and it shows no $S_8$ tension, but a remnant $\sim3.5\sigma$ Hubble tension \cite{Motloch:2018pjy,Motloch:2019gux}. Interestingly, the cosmology deduced once marginalizing over the lensing information is in better agreement with the recent results from the SPTPol \cite{Aghanim:2016sns,Henning_2018,Chudaykin:2020acu}, which shows no tension with the LSS measurement of $S_8$, a weaker $H_0$ tension, and no lensing anomaly. Pin-pointing the source of such lensing anomaly (perhaps a simple statistical fluke, although quantifying is likelihood and how to treat it is complicated) is therefore of utmost importance to understand whether the $S_8$ tension derives from it.
 
Motivated by this fact, we perform an analysis of the $\Lambda$CDM and EDE cosmology against {\it Planck} and a prior on $S_8$ from the joint DES-Y1 and KiDS results, while marginalizing over the lensing information. We find that both the unlensed $\Lambda$CDM and EDE cosmology spectrum agrees better with LSS data, and that the presence of EDE does not affect the amount of anomalous lensing. This means that the anomalous lensing is not due to the EDE, and also that the success of EDE is not due to opening up a new degeneracy direction with some exotic lensing parameters.

The paper is structured as follows. In sec.~\ref{sec:EDE}, we present the phenomenological EDE models studied in this work and compare it to other models from the literature. In sec.~\ref{sec:Planck}, we perform an `anatomy' of the EDE resolution to the Hubble tension, to understand how each data set reacts to the presence of EDE. We additionally introduce the new baseline 1-parameter EDE model that is favored by \Planck~2018 data. We present our $N$-Body simulations validating the HMcode prediction and confront the 1-parameter EDE model against WL data in sec.~\ref{sec:confront}. We discuss the $H_0$ and $S_8$ tension in light of the lensing-marginalized CMB spectrum in sec.~\ref{sec:towards} and suggest some promising ways of restoring cosmological concordance in an extended EDE cosmology. Finally, we conclude in sec.~\ref{sec:concl}.

\section{The phenomenological Early Dark Energy model resolving the Hubble Tension}
\label{sec:EDE}

The possible presence of a dark energy component before last-scattering has been studied for more than a decade \cite{Doran:2000jt,Wetterich:2004pv}. These alternative cosmological realizations have little to do with that under study here, as they typically assume tracking equation of state at early times. The idea of an anomalous era of expansion triggered by a frozen scalar field as a resolution to the Hubble tension was introduced in Ref.~\cite{Karwal:2016vyq}, where a background-only computation was shown to alleviate the Hubble tension. However, it is the work of Ref.~\cite{Poulin:2018cxd} that showed through a fluid approximation the key role played by perturbations in the scalar field to allow for a resolution of the Hubble tension. Since this work, the treatment of the EDE component has been improved \cite{Agrawal:2019lmo,Smith:2019ihp,Lin:2019qug}, and augmented to deal with alternative potentials and better motivated underlying fundamental models \cite{Kaloper:2019lpl,Agrawal:2019lmo,Lin:2019qug,Sakstein:2019fmf,Berghaus:2019cls,Alexander:2019rsc,Braglia:2020bym,Ballesteros:2020sik,Gonzalez:2020fdy,Ballardini:2020iws,Niedermann:2019olb,Niedermann:2020dwg}. 
In particular, it has been shown that {\it Planck} data not only provide a detection of the background dynamics of the EDE component, but also severely restricts the dynamics of perturbations \cite{Lin:2019qug,Smith:2019ihp}. As such, {\it Planck} data allows for pinning down directly properties of the EDE, making the choice of model crucial. They favor either non-canonical kinetic term whereby the equation of state $w$ is approximately equal to the effective sound speed $c_s^2$ \cite{Lin:2019qug}, or potential that flattens close to the initial field value \cite{Smith:2019ihp}.

In this work, we study the modified axion potential introduced in Refs.~\cite{Kamionkowski:2014zda,Karwal:2016vyq,Poulin:2018dzj,Poulin:2018cxd,Smith:2019ihp},
\begin{equation}\label{eq:potential}
    V_n(\Theta) = m^2 f^2[1-\cos (\Theta)]^n,
\end{equation} 
where $m$ represent the axion mass, $f$ the decay constant and $\Theta \equiv \phi/f$ is a re-normalized field variable, so that $-\pi \leq \Theta \leq \pi$. We assume that the field always starts in slow-roll the background dynamics and without loss of generality we restrict $0\leq \Theta_i \leq \pi$.

This potential is a phenomenological generalization of the well motivated axion-like potential (which can be recovered by setting $n=1$) that arise generically in string theory \cite{Svrcek:2006yi,Douglas:2006es,Arvanitaki:2009fg,Marsh:2015xka,Marsh:2015xka}. 
Such a potential may be generated by higher-order instanton corrections \cite{Kappl:2015esy}, but taken at face values would suffer from a strong fine-tuning issues necessary to the cancelling of the lowest orders instantons. Therefore, it should not be interpreted beyond a phenomenological description. 
We note that similar forms of potential, with power law minima and flattened ``wings" have been used in the context of inflationary physics, as well as dark energy (see, e.g.,~Refs.~\cite{Dong:2010in,Kallosh:2013hoa,Carrasco:2015pla,Geng:2019phi}).
Still, this form was devised to allow for flexibility in the background dynamics after the field becomes dynamical, and it also provides an excellent fit to both {\it Planck} and SH0ES data. In fact, to the best of our knowledge, it corresponds to the EDE scenario that leads to the best combined $\chi^2$ of the cosmological data-sets under study (although the better theoretically motivated model studied in Ref.~\cite{Niedermann:2019olb,Niedermann:2020dwg} seems to perform equally well). 

\begin{figure}
    \centering
    \includegraphics[scale=0.43]{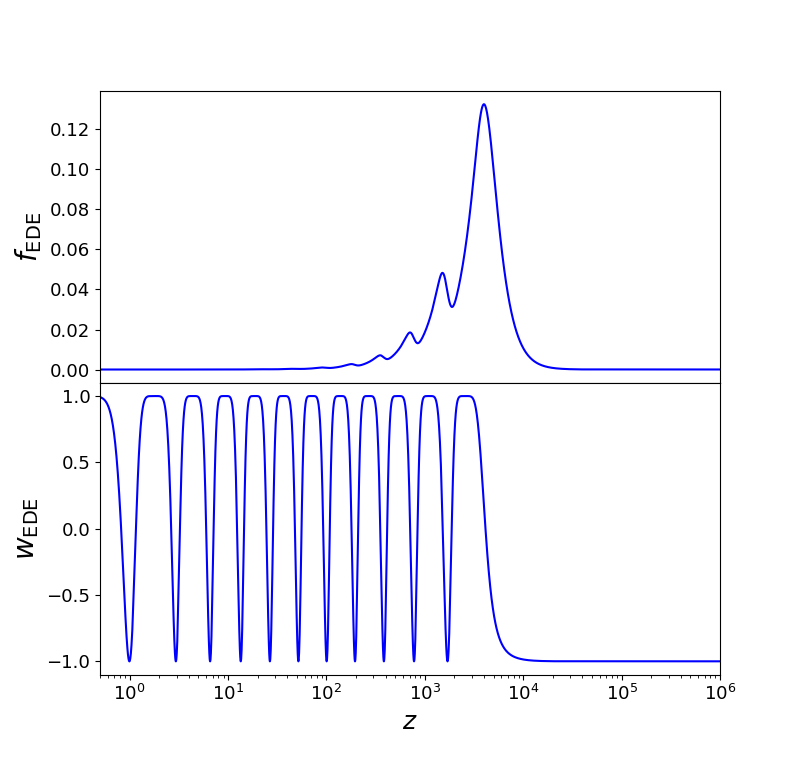}
    \caption{Fractional energy density (upper) and equation of state (lower) in the EDE as a function of redshift. Cosmological parameters are set to the best-fit values from the \Planck+\BAO+SNIa+\SHOES~analysis (see third column of Table \ref{table:param_values_3par}).}
    \label{fig:f_and_w_EDE}
\end{figure}

We refer to Refs.~\cite{Poulin:2018dzj,Smith:2019ihp} for all necessary details about the model. 
However, the key features can be summarized as follows: at early times the scalar field is frozen due to Hubble friction, until the Hubble rate drops below its mass value; the field then starts moving in the potential, and eventually oscillating around the minimum, at which point the energy density dilutes at a rate dictated by the asymptotic equation of state $w(n) = (n-1)/(n+1)$ (e.g., Refs.~\cite{Griest:2002cu,Marsh:2010wq,Poulin:2018dzj}).
In Fig. \ref{fig:f_and_w_EDE} we show the redshift evolution for the fractional energy density and the equation of state in the EDE. 

We can trade three out of the four model parameters $\{m,f,n,\Theta_i\}$ for phenomenological parameters: the first two of them describing the fractional energy density $f_{\rm EDE}(z_c)$ at the critical redshift $z_c$ where the field becomes dynamical and the asymptotic equation of states after the field becomes dynamical $w(n) = (n-1)/(n+1)$, respectively; the last degree of freedom lies in the dynamics of linear perturbations, whose phenomenology is captured by the effective sound speed $c_s^2$. However, within the EDE scalar field scenario under study, such freedom is intrinsically encoded in the choice of the initial field value\footnote{In practice, it is the curvature of the potential, $\partial^2 V(\Theta)/\partial^2\Theta$, close to the initial field value $\Theta_i$ that dictates the last of degree of freedom in the perturbation dynamics \cite{Poulin:2018dzj,Smith:2019ihp}.} $\Theta_i$, once the other phenomenological parameters have been fixed. 

To perform our analyses, we use the modified version of the Einstein-Boltzmann code {\sc CLASS} \cite{Lesgourgues:2011re,Blas:2011rf} presented in Ref.~\cite{Smith:2019ihp}. The code is publicly available at \url{https://github.com/PoulinV/AxiCLASS} (the latest version, used for this study, can be found in the ``devel'' branch).

\section{Confronting EDE to Planck 2018, BOSS and Pantheon data}
\label{sec:Planck}
In this section, we test the EDE scenario with various combinations of data-sets, in order to extract the cosmology that would resolve the Hubble tension, and compare with results from past literature making use of Planck 2015 data.
We will test our phenomenological model against:

\begin{itemize}
    \item \PlanckTTTEEELensing: the high-$\ell$ TT,TE,EE, low-$\ell$ TT and EE data from {\em Planck} 2018 through the baseline {\sc plik}, {\sc commander} and {\sc simall} likelihoods \cite{Aghanim:2019ame,Aghanim:2018oex}, alone and combined with the lensing amplitude reconstruction ({\sc simca} likelihood); we make use of a Cholesky decomposition as implemented in {\sc MontePython-v3} to handle the large number of nuisance parameters \cite{Lewis:2013hha}.
    \item \BAO: the measurements from 6dFGS at $z = 0.106$ \cite{Beutler:2011hx}, SDSS MGS at $z = 0.15$ \cite{Ross:2014qpa}, and BOSS DR12 data at $z = $ 0.38, 0.51 and 0.61 \cite{Alam:2016hwk}.
    \item \FS: the measurements of the growth function $f\sigma_8(z)$ (FS)  from the CMASS and LOWZ galaxy samples of BOSS DR12 at $z = 0.38$, $0.51$, and $0.61$~\cite{Alam:2016hwk}. In practice, we make use of the ``concensus'' BAO and FS result that combines both in a single likelihood\footnote{We correct for a small mistake in the standard \textsc{Montepython-v3} implementation of the likelihood. This also explains why Refs.~\cite{Poulin:2018cxd,Smith:2019ihp}, which did not correct for this mistake, reported slightly different constraints when including this likelihood.}.
    \item \Pantheon: the Pantheon SNIa catalogue, spanning redshifts $0.01 < z < 2.3$; we marginalize over the nuisance parameter $\cal{M}$ describing the SNIa calibration.
    \item \SHOES: the SH0ES result, modelled with a Gaussian likelihood centered on $H_0 = 74.03 \pm 1.42$ km/s/Mpc \cite{Riess:2019cxk}; however, choosing a different value that {\it combines} various direct measurements would not affect the result, given their small differences. 
  
\end{itemize}
\subsection{Baseline analysis: anatomy of the 3-parameter EDE model resolving the Hubble tension}

\begin{figure*}[t]
    \centering
\includegraphics[scale=0.52]{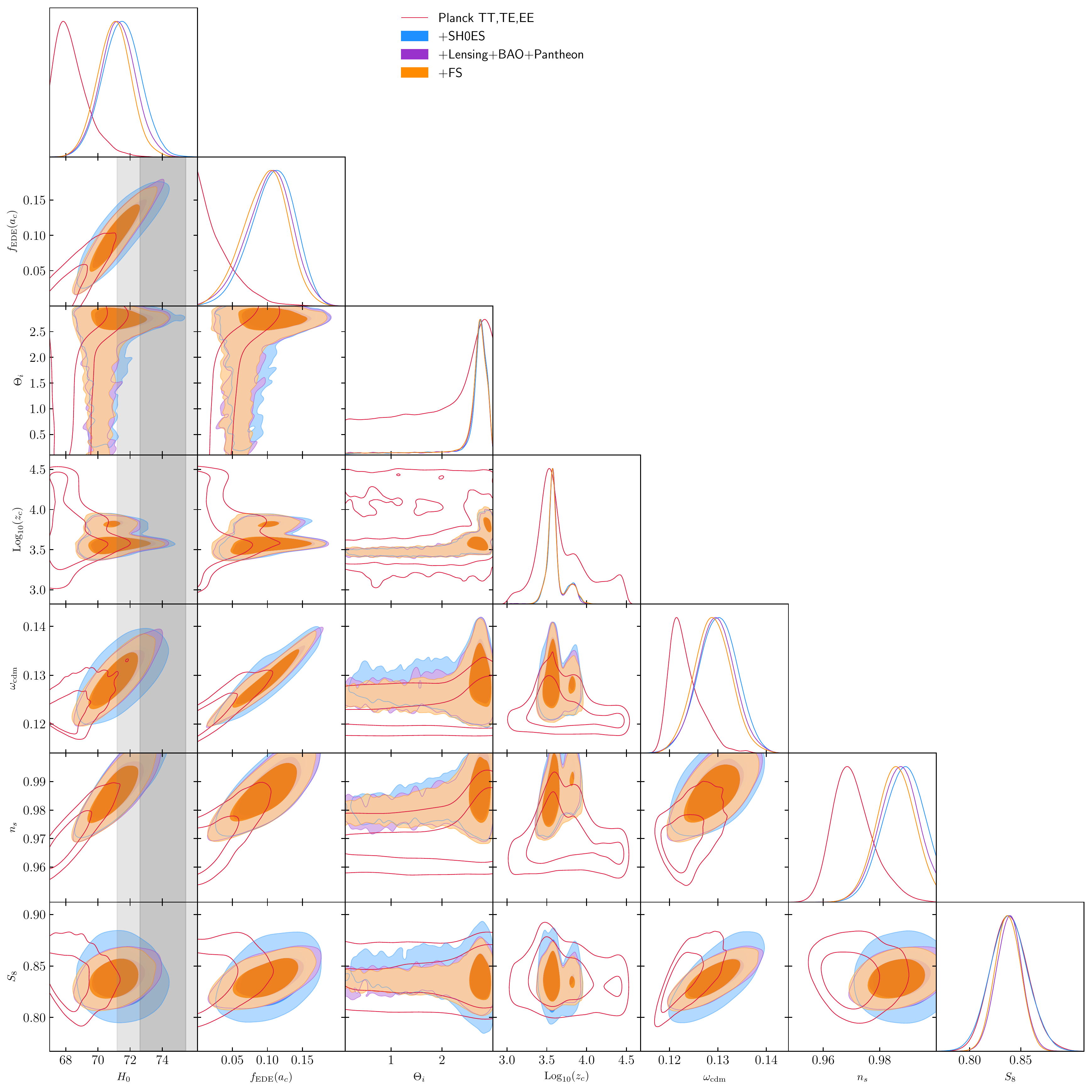}
    \caption{Reconstructed 2D posterior distributions of a subset of parameters for various data set combinations (see legend) in the 3-parameters EDE cosmology. }
    \label{fig:posterior-3param}
\end{figure*}

Our baseline cosmology consists in the following combination of the six $\Lambda$CDM parameters $\{\omega_b,\omega_{\rm cdm},H_0,n_s,A_s,\tau_{\rm reio}\}$, plus three parameters describing the EDE sector, namely $\{f_{\rm EDE}(a_c),{\rm Log}_{10}(a_c),\Theta_i\}$. We use wide flat priors on all these parameters. We follow the Planck convention and assume two massless neutrinos and one massive with $m_\nu=0.06$ eV. We perform our MCMC analyses using {\sc MontePython-v3}, and consider chains to be converged with the Gelman-Rubin criterion\footnote{Most chains are in fact converged at the $R-1\sim0.01$ level, this somewhat `loose' but reasonable criterion was only used once including \KV~data,  which are much longer to converge.} $R-1<0.1$ \cite{Gelman:1992zz}. To extract best-fit parameters, we make use of the {\sc Minuit} algorithm \cite{James:1975dr} through the {\sc iMinuit} python package\footnote{\url{https://iminuit.readthedocs.io/}}.
Starting from Planck only, we now discuss the impact of adding data-sets on the reconstructed EDE parameters. We compare the evolution of the  $\chi^2$ in the EDE cosmology as we add data-sets, to that of the $\Lambda$CDM model in the same combined fit, with and without \SHOES~data.
The results are presented in Tab.~\ref{table:param_values_3par} and we show the 1D and 2D posterior distributions of $\{H_0,f_{\rm EDE}(z_c),\Theta_i,{\rm Log}_{10}(z_c),\omega_{\rm cdm},n_s,S_8\}$ in Fig.~\ref{fig:posterior-3param}. All relevant $\chi^2$ information is given in App.~\ref{sec:Appendix_chi2}.

\begin{center}
\begin{table*}[t!]
\scalebox{0.9}{
  \begin{tabular}{|l|c|c|c|c|}
     \hline
     \multicolumn{5}{|c|}{3-parameter EDE cosmology} \\
     \hline Parameter &\PlanckTTTEEE&+\SHOES&+\PlanckLensing+\BAO+\Pantheon&+\FS\\ \hline \hline
    $H_0$ [km/s/Mpc] &  $68.29(70.49)_{-1.3}^{+0.75}$ & $71.49(73.05) \pm 1.2$ &$71.34(72.41)_{-1.1}^{+1}$ &  $71.01(71.96)_{-1}^{+1.1}$\\
    $100~\omega_b$  & $2.252(2.270)_{-0.023}^{+0.019}$& $2.284(2.281)_{-0.024}^{+0.022}$ & $2.282(2.292)_{-0.022}^{+0.021}$&  $2.28(2.285)_{-0.022}^{+0.021}$ \\
    $\omega_{\rm cdm}$&  $0.1232(0.1278)_{-0.004}^{+0.0019}$ &$0.13(0.135)_{-0.004}^{+0.0042}$ &  $0.1297(0.1327)_{-0.0039}^{+0.0036}$&  $0.1289( 0.1323) \pm 0.0039$\\
    $10^{9}A_s$& $2.116(2.124)_{-0.041}^{+0.035}$& $2.153(2.160)_{-0.042}^{+0.036}$&$2.152(2.183)_{-0.035}^{+0.031}$ &  $2.144(2.135)_{-0.033}^{+0.032}$  \\
    $n_s$& $0.9706(0.9829)_{-0.0087}^{+0.0058}$&$0.9889(0.9966)_{-0.0075}^{+0.0076}$ & $0.9878(0.9963)_{-0.007}^{+0.0066}$ & $0.9859(0.9895)_{-0.0071}^{+0.007}$  \\
    $\tau_{\rm reio}$& $0.0552(0.0524)_{-0.0086}^{+0.0076}$ &  $0.0586(0.0558)_{-0.0091}^{+0.0077}$ & $0.0585(0.0633)_{-0.008}^{+0.007}$&$0.0574(0.0528)_{-0.0079}^{+0.007}$  \\
    $f_{\rm EDE}(z_c)$ & $< 0.088 (0.085)$& $0.108(0.152)_{-0.028}^{+0.035}$ &$0.106(0.133)_{-0.028}^{+0.031}$ & $0.097(0.126)_{-0.029}^{+0.035}$  \\
    ${\rm Log}_{10}(z_c)$& $3.705(3.569)_{-0.22}^{+0.37}$ &  $3.612(3.569)_{-0.049}^{+0.13}$& $3.615(3.602)_{-0.029}^{+0.11}$&$3.61(3.572)_{-0.054}^{+0.13}$  \\
    $\Theta_i$ & unconstrained (2.775)&  $2.604(2.756)_{0.0087}^{+0.33}$&  $2.722(2.759)_{-0.092}^{+0.17}$ & $2.557(2.705)_{0.025}^{+0.37}$   \\
    \hline
    $100~\theta_s$ & $1.04165(1.04371)_{-0.00034}^{+0.00039}$&   $1.04131(1.04070)_{-0.0004}^{+0.00039}$& $1.04143(1.04122)_{-0.00039}^{+0.00036}$&  $1.04145(1.04098)_{-0.00039}^{+0.00038}$\\
    $r_s(z_{\rm rec})$& $142.8(140.1)_{-0.72}^{+1.9}$&$138.8(136.4)_{-1.9}^{+1.7}$ & $139(137.5)_{-1.7}^{+1.7}$&$139.4(137.8)_{-1.9}^{+1.7}$\\
    $S_8$ & $0.839(0.834)_{-0.019}^{+0.018}$ &$0.838(0.842)_{-0.019}^{+0.018}$ &  $0.838(0.846 ) \pm 0.013$ & $0.837(0.838) \pm 0.013$ \\
    $\Omega_m$ & $0.314(0.304)_{-0.0091}^{+0.0088}$ &  $0.3004(0.2969)_{-0.0084}^{+0.0079}$& $0.301(0.2980)_{-0.0055}^{+0.0051}$ & $0.3022(0.3009)_{-0.0054}^{+0.0053}$\\
    \hline
    $\Delta \chi^2_{\rm min}$ ($\Lambda$CDM w/ SH0ES) & $-$ & -20.8& -19.1  &-18.7\\
    $\Delta \chi^2_{\rm min}$($\Lambda$CDM w/o SH0ES) & -4.9& -1.5& -0.02 &-0.6\\
    \hline
  \end{tabular}}
  \caption{The mean (best-fit) $\pm1\sigma$ error of the cosmological parameters reconstructed from the combined analysis of various data sets (from left to right, each column adds a set of data to the previous one). We also report the $\Delta\chi^2_{\rm min}$ with respect to a $\Lambda$CDM fit to the same data-sets, with and without a prior on $H_0$ from \SHOES.}
\label{table:param_values_3par}
\end{table*}
\end{center}

\vspace{0.1cm}

{\bf Planck TT,TE,EE~only:} with \PlanckTTTEEE~data only and three free parameters, the EDE model under study is not detected. In agreement with Refs.~\cite{Hill:2020osr, Niedermann:2020dwg}, we find that the fraction of EDE at $z_c$ is limited to\footnote{Hereinafter, we quote 1-sided constraints at 95\%C.L., and two-sided ones at 68\%C.L.} $f_{\rm EDE}(z_c)<0.088$, while \Logzc~and $\Theta_i$ are unconstrained. Interestingly, we also find that the best fit within \Planck~data only has $\{$\fEDE~$\sim8.5\%$, \Logzc~$\sim 3.56$, $\Theta_i\sim2.8,H_0\sim70.5$~km/s/Mpc$\}$ and a $\Delta\chi^2_{\rm min}\equiv\chi^2_{\rm min}(\Lambda{\rm CDM})-\chi^2({\rm EDE}) \simeq-5$ in favor of the EDE model\footnote{To guide the reader, we mention that a  1$\sigma$ shift in the quality of the fit to \Planck~data roughly corresponds to a $\Delta\chi^2$ of $\sim6$ (see the distribution of \Planck's $\chi^2$ in the tables available \href{https://wiki.cosmos.esa.int/planck-legacy-archive/images/b/be/Baseline_params_table_2018_68pc.pdf}{at this link}). }. One can already note a curiosity: the best fit value of \fEDE~is very close to the $2\sigma$ bound that we obtain. This, as we will discuss later, is due to the choice of flat, uninformative prior on \Logzc~and $\Theta_i$.

\vspace{0.1cm}

{\bf Planck TT,TE,EE+SH0ES:} Once a prior on $H_0$ given by \SHOES~is included in the analysis, the sampler explores more easily a part of parameter space with higher $H_0$ values, and the EDE is now well detected: $\{$\fEDE$\simeq0.11^{+0.036}_{-0.031}$, \Logzc$= 3.6_{-0.039}^{+0.14}$, $\Theta_i=2.569_{-0.032}^{+0.36}\}$, with $H_0 = 71.4\pm1$ km/s/Mpc.  This is in excellent agreement with results from past literature \cite{Poulin:2018cxd,Lin:2019qug,Smith:2019ihp,Niedermann:2020dwg}. Remarkably, the best fit values of both $\Theta_i$ and \Logzc~ are in perfect agreement with that obtained \Planck~only. This is highly non-trivial, and seem to indicate that \Planck~does favor the region of the \{\Logzc,$\Theta_i$\}-space that resolves the Hubble tension.  However, the best fit fraction reaches $15\%$, a value that one would naively consider to be strongly excluded by the \Planck~only analysis. In fact, that is not the case, as the the fit to \Planck~data is barely affected by the additional $H_0$ prior, while one can get a perfect fit of \SHOES~data. Concretely, the $\chi^2_{\rm min}({\rm EDE})$ when fitting \Planck+SH0ES increases by~$\sim3$, such that even in this combined fit, the $\chi^2$ of \Planck~data is smaller than that of $\Lambda$CDM fitted on \Planck~data only. This indicates that the limit on \fEDE~derived in a \Planck~only analysis is not robust, as it is entirely driven by our choice of flat priors. This was also discussed in Refs.~\cite{Smith:2019ihp,Niedermann:2020dwg}, and the reason for that is clear\footnote{Here, let us mention that Ref.~\cite{Ivanov:2020ril} make the comment that such degeneracy does not exist. This is of course only true because they include the $3\sigma$ discrepant $S_8$ data to their analysis. The degeneracy is very clear within \Planck~data.}: there exists a strong $\chi^2$ degeneracy in \Planck~data between $\Lambda$CDM and the EDE cosmology, that, given our choice of uninformative priors on $\Theta_i$ and \Logzc, leads to an artificially strong bound on \fEDE. Indeed, once \fEDE~drops below $\lesssim 4\%$ (as seen from the 2D posterior), its impacts on the power spectrum is not detectable given current measurement accuracy. As a result, the quantity \Logzc
~and $\Theta_i$ have no impact on observables, such that any choice of \Logzc~and $\Theta_i$ leads to a cosmology indistinguishable from $\Lambda$CDM. Therefore, the sampler spends much more time exploring this degeneracy direction, rather than efficiently sampling the narrow degeneracy between \fEDE~and $H_0$, which requires a specific choice of \Logzc~and $\Theta_i$ to appear. Following Ref.~\cite{Niedermann:2020dwg}, we will discuss a natural way to alleviate this issue in Section~\ref{sec:1param}.

\vspace{0.1cm}

{\bf Planck TT,TE,EE+PP+BAO+SNIa+SH0ES:} We now add to our analysis the lensing reconstruction \PlanckLensing, the \Pantheon~SNIa data-set, and the \BAO~data from BOSS. Strikingly, the addition of these three data-sets has almost no impact on the reconstructed posteriors, nor on the best fit. This is far from a trivial test to pass, as many of the suggested resolutions to the Hubble tension are strongly constrained by the addition of these data-sets \citep{Beutler:2011hx,Ross:2014qpa,Riess:2016jrr,DiValentino:2017zyq,Addison:2017fdm,DiValentino:2017iww,Alam:2016hwk,Bernal:2016gxb,Zhao:2017cud,Poulin:2018zxs,Aylor:2018drw,2020PhRvD.101h3524R}. However, as noted in Refs.~\cite{Poulin:2018cxd,Smith:2019ihp,Niedermann:2020dwg}, we find that the reconstructed $\omega_{\rm cdm}$ and $n_s$ in the EDE cosmology are somewhat higher than in $\Lambda$CDM, such that the $S_8$ tension is slightly increased. As suggested in past literature \cite{Hill:2020osr,Ivanov:2020ril,DAmico:2020ods}, this opens up the possibility of constraining the EDE resolution using LSS data, and in particular the $S_8$ measurement from weak gravitational lensing surveys.\looseness=-1~However, combining KiDS+VIKING/HSC data with \Planck~to constrain extension to $\Lambda$CDM can be problematic as: i) they require the ability to predict the non-linear power spectrum at relatively small scales in models beyong $\Lambda$CDM; ii) the $\Lambda$CDM best fit model from \Planck~is not a good fit to these data.

\vspace{0.1cm}

{\bf All Data:} As a starter, we add the `consensus' $f\sigma_8$ BOSS likelihood to the analysis, which is consistent with the $\Lambda$CDM model from \Planck; we find a mild $\sim0.4\sigma$ decrease in the reconstructed mean, now being $H_0 \simeq 71\pm1$ km/s/Mpc and $f_{\rm EDE}\simeq 0.1\pm0.03$. This is consistent with the fact that the $f\sigma_8$ measurements are sensibly lower than the $\Lambda$CDM prediction, while the EDE cosmology leads to slightly larger values. Still, the $\chi^2$ of the \FS~data is barely affected; in fact, as before, $\Lambda$CDM provides a slightly worse fit to the joint data-set, even when the \SHOES~prior is not included in the analysis. Before including weak lensing measurements to the analysis, we therefore conclude that the 3-parameter EDE model under study performs very well in resolving the Hubble tension, but future measurement of $f\sigma_8$ will certainly put the model under crucial tests.

\subsection{Towards a 1-parameter resolution to the Hubble tension}
\label{sec:1param}

\begin{figure*}
    \centering
        \includegraphics[scale=0.7]{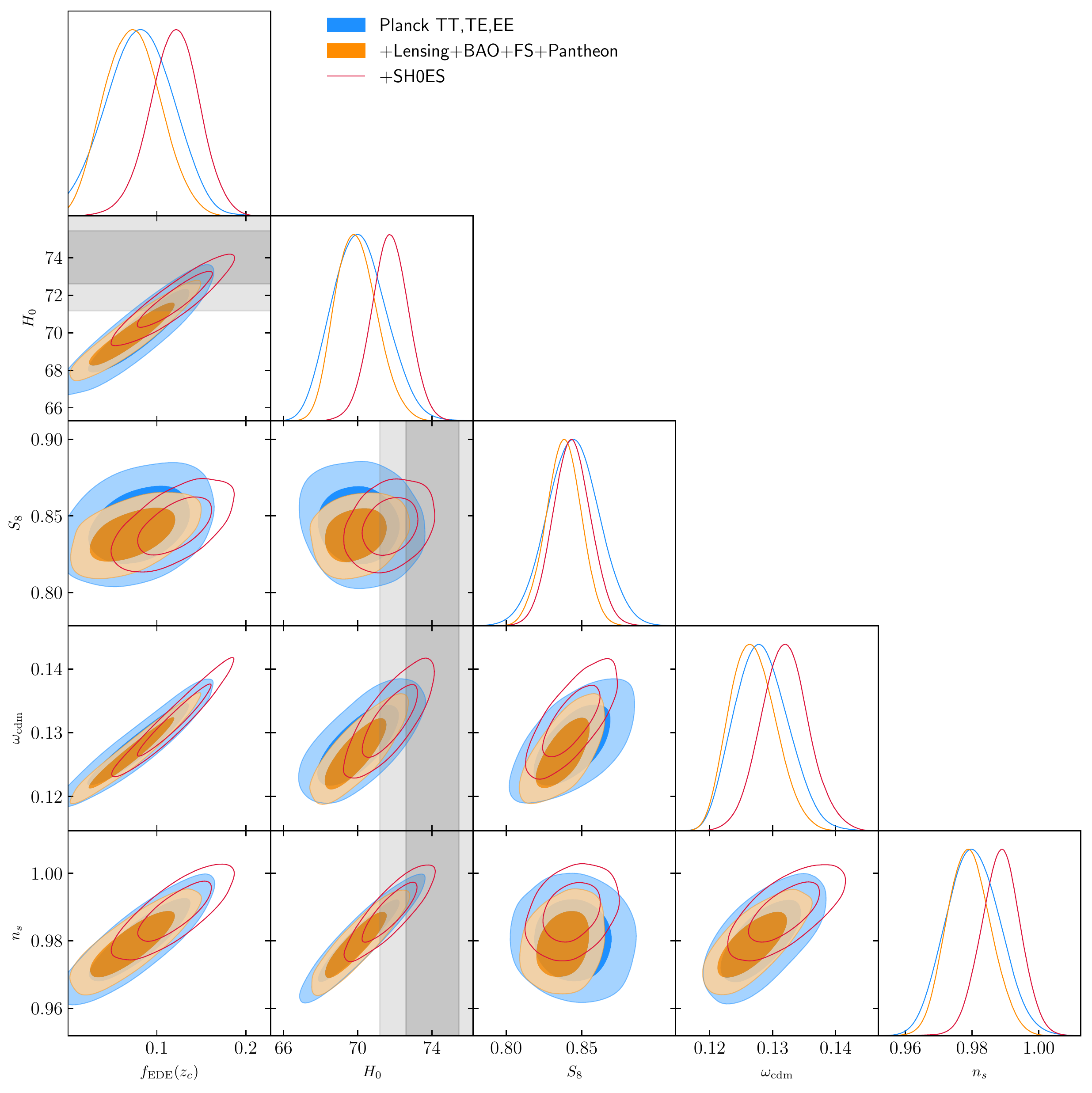}

    \caption{Reconstructed 2D posterior distributions of a subset of parameters for various data set combinations (see legend) in the 1-parameter EDE cosmology. }
    \label{fig:posterior-1param}
\end{figure*}

\begin{table*}[!t]
\scalebox{0.9}{
  \begin{tabular}{|l|c|c|c|}
     \hline
     \multicolumn{4}{|c|}{1-parameter EDE cosmology} \\

     \hline Parameter &\PlanckTTTEEE&+\PlanckLensing+\BAO+\Pantheon+\FS&+\SHOES\\ \hline \hline
    $H_0$ [km/s/Mpc] &  $70.10(70.83)_{-1.6}^{+1.4}$  &  $70.00(69.84)_{-1.4}^{+0.98}$ &  $71.71 (72.21)_{-0.95}^{+1.0}$ \\
    $100~\omega_b$  &  $2.258(2.265)_{-0.018}^{+0.018}$ & $2.259(2.263)_{-0.016}^{+0.015}$  &  $2.273(2.282) \pm 0.013$\\
    $\omega_{\rm cdm}$& $0.1282(0.1306)_{-0.0046}^{+0.0039}$  & $0.1270(0.1265)_{-0.0043}^{+0.0033}$ & $0.1317(0.1310)_{-0.0037}^{+0.0036}$ \\
    $10^9A_s$ & $2.137(2.164)_{-0.042}^{+0.037}$ & $2.131(2.118)_{-0.036}^{+0.03}$& $2.15(2.140)_{-0.031}^{+0.032}$\\
    $n_s$&$0.9803(0.9851)_{-0.0085}^{+0.0079}$ & $0.9795(0.9788)_{-0.0074}^{+0.0063}$ & $0.9884(0.9917)_{-0.0057}^{+0.0062}$  \\
    $\tau_{\rm reio}$& $0.0558(0.0602)_{-0.0085}^{+0.0079}$&  $0.0553(0.0522)_{-0.0075}^{+0.0069}$ & $0.0561(0.0536)_{-0.0076}^{+0.0071}$ \\
    $f_{\rm EDE}(z_c)$ &$0.082(0.104)_{-0.038}^{+0.037}$ &$0.074(0.070)_{-0.036}^{+0.03}$  &  $0.118(0.122)_{-0.026}^{+0.029}$\\
    \hline
    $100~\theta_s$ &$1.04147(1.0413)_{-0.00035}^{+0.00036}$ & $1.04153(1.04159)_{-0.00032}^{+0.00035}$ & $1.04157(1.04127) \pm 0.00034$\\
    $r_s(z_{\rm rec})$&  $140.1(138.9)_{-2.0}^{+2.2}$ & $140.6(140.8)_{-1.6}^{+2.1}$ &  $138.2(138.2)_{-1.8}^{+1.6}$ \\
    $S_8$ & $0.844(0.851)_{-0.018}^{+0.017}$  & $0.838(0.835) \pm 0.012$   & $0.843(0.839)_{-0.013}^{+0.012}$ \\
    $\Omega_m$ & $0.3084(0.3067)_{-0.0093}^{+0.009}$& $0.3067(0.3071)_{-0.0058}^{+0.0055}$&  $0.3017(0.2962)_{-0.0054}^{+0.0051}$\\
    \hline
    $\Delta \chi^2_{\rm min}$ ($\Lambda$CDM) & -5 & -6 & -18.5  (-0.5)  \\
    \hline
  \end{tabular}
  }
  \caption{The mean (best-fit) $\pm1\sigma$ error of the cosmological parameters reconstructed from the combined analysis of various data-sets  (from left to right, each column adds a sets of data to the previous one). We also report the $\Delta\chi^2_{\rm min}$ with respect to a $\Lambda$CDM fit to the same data-sets. In the last row, we also report the $\Delta\chi^2$ with respect to $\Lambda$CDM fit to the combined data without SH0ES in parenthesis.}
  \label{table:param_values_1par}
\end{table*}

Before turning to the inclusion of WL data, we show that the apparently tight bounds obtained when the \SHOES~prior on $H_0$ is not included is due to our choice of uninformative priors for $\Theta_i$ \& \Logzc. In fact, one can strongly weaken the bound on $f_{\rm EDE}$ by {\em reducing} the EDE parameter space to a suitable choice of \Logzc~and $\Theta_i$. This might sound counter-intuitive: in principle, one expects to relax constraints on a given parameter by enlarging the parameter space such as to introduce a new degeneracy. Here however, it is the poor prior choice which leads to a strong bound on \fEDE~independently of the data combination. Fixing \Logzc~and $\Theta_i$ to some fiducial values surely rises the question of what values should one choose. In a realistic scenario, one might know these values a priori; one example is the scenario discussed in Ref.~\cite{Sakstein:2019fmf} in which a scalar field experiences a phase-transition around the redshift at which neutrinos becomes non-relativistic, such that the critical redshift is specified by the value of the neutrino mass, while $\Theta_i$ is set by the dynamics of the phase-transition (see also Refs.~\cite{Berghaus:2019cls,Braglia:2020iik,Ballesteros:2020sik,Gonzalez:2020fdy,Ballardini:2020iws,Niedermann:2020dwg} for different EDE models with fewer free parameters).  Here however, we have been considering a phenomenological model whose primary characteristics is to have enough freedom to extract information from the data to resolve the tension -- we will therefore make use of that information and fix $\Theta_i$ \& \Logzc~to their best fit value from \Planck~data only -- which, we recall, are close-to-identical to that obtained in the combined fit.  
We report in Table~\ref{table:param_values_1par} the reconstructed cosmological parameters from \Planck~only and from the combined fit of all data, with and without including \SHOES. We show the reconstructed 2D posteriors of $\{f_{\rm EDE}(z_c), H_0,S_8\}$ in Figure~\ref{fig:posterior-1param}. Notice how the degeneracy direction \fEDE$-H_0$ clearly opens up. Furthermore, the mild $\Delta\chi^2$ preference in favor of the EDE cosmology now leads to reconstructing \fEDE$=0.082\pm0.037$, i.e., a $\sim 2\sigma$ preference for non-zero EDE {\em from \Planck~data only}. The inferred $H_0=70.1\pm1.4$km/s/Mpc is now in agreement with the SH0ES determination at better than\footnote{{From here on, we quote `tension' and `agreement' assuming Gaussian posteriors for simplicity. While this is surely a crude approximation (to be tested elsewhere \cite{Karwal2021} following Ref.~\cite{Raveri:2018wln}), we believe this is justified because the posterior of interest $(H_0,S_8)$ are close to Gaussian, and since this is the approach followed by collaborations when quoting tensions in the $\Lambda$CDM context \cite{Riess:2019cxk,Aghanim:2018eyx}.}} $2\sigma$. The addition of \BAO,~\FS~and \Pantheon~measurements has little impact; the reconstructed EDE fraction shifts downward by $\sim0.3\sigma$, slightly degrading the success of the resolution to the Hubble tension, while the $2\sigma$ preference for non-zero EDE is still present. These results are in excellent agreement with these presented in Ref.~\cite{Niedermann:2020dwg} for a different EDE model.
Finally, the inclusion of a prior from \SHOES~pulls up the fraction of EDE to \fEDE$=0.118\pm0.029$ and the value of $H_0 = 71.7\pm1$ km/s/Mpc, at the cost of a small degradation in $\chi^2_{\rm min}$ ($\Delta\chi^2\sim+6$). Yet, as before, the $\chi^2_{\rm min}$ of the {\em combined} fit \Planck+\BAO+\FS+\Pantheon+\SHOES~in the 1-parameter EDE cosmology is slightly lower than a $\Lambda$CDM fit to \Planck+\BAO+\FS+\Pantheon~(no \SHOES). This attests that, despite this small degradation in $\chi^2_{\rm min}$, the goodness of fit is still excellent.
However, as discussed previously, the values of $S_8$ are in significant tension with weak lensing measurements, and one might expect that it is possible to strongly constrain the EDE model by including LSS data. We study this possibility in detail in Section~\ref{sec:confront}, following Refs.~\cite{Hill:2020osr,DAmico:2020ods,Ivanov:2020ril}.

\section{Confronting EDE to weak lensing surveys}\label{sec:confront}
In order to make use of weak gravitational lensing data to perform LSS analyses, one needs to accurately model the matter power spectrum in the late-time non-linear regime. To this purpose, one can adopt the {\sc{halofit}} semi-analytical prescription \cite{Smith:2002dz}, as revised by \cite{Takahashi:2012em}, which has been shown to be accurate at 5$\%$ level in reproducing the non-linear power spectra of $\Lambda$CDM models up to wavenumbers $k \leq 10 \, h/{\rm Mpc}$. However, the version developed by authors of Ref.~\cite{Takahashi:2012em} does not consider the impact of baryon feedback. A further improvement, dubbed as {\sc{HMcode}}, has been developed in Ref.~\cite{Mead:2015yca}, its main advantage being its flexibility to account for the effects of baryon physics on the small-scale clustering of matter, particularly important at very low redshifts. 
Both {\sc{halofit}} and {\sc{HMcode}} have been shown to be suitable to describe the $\Lambda$CDM scenario, as well as some common extensions beyond it, such as models with varying DE EoS or massive neutrinos \cite{Joudaki:2016kym}. In Section \ref{sec:sims} we confront the non-linear matter power spectra produced by using {\sc{halofit}}/{\sc{HMcode}} in the EDE framework, against the outputs of dedicated cosmological $N$-Body simulations, to explicitly demonstrate the accuracy of our LSS data analyses. In Section~\ref{subsec:mcmc} we discuss the results of our MCMC analysis against weak lensing data.

\subsection{Non-linear matter power spectrum: a comparison with $N$-Body simulations}\label{sec:sims}

\begin{figure*}
\begin{tabular}{cc}
\includegraphics[width=.499\textwidth]{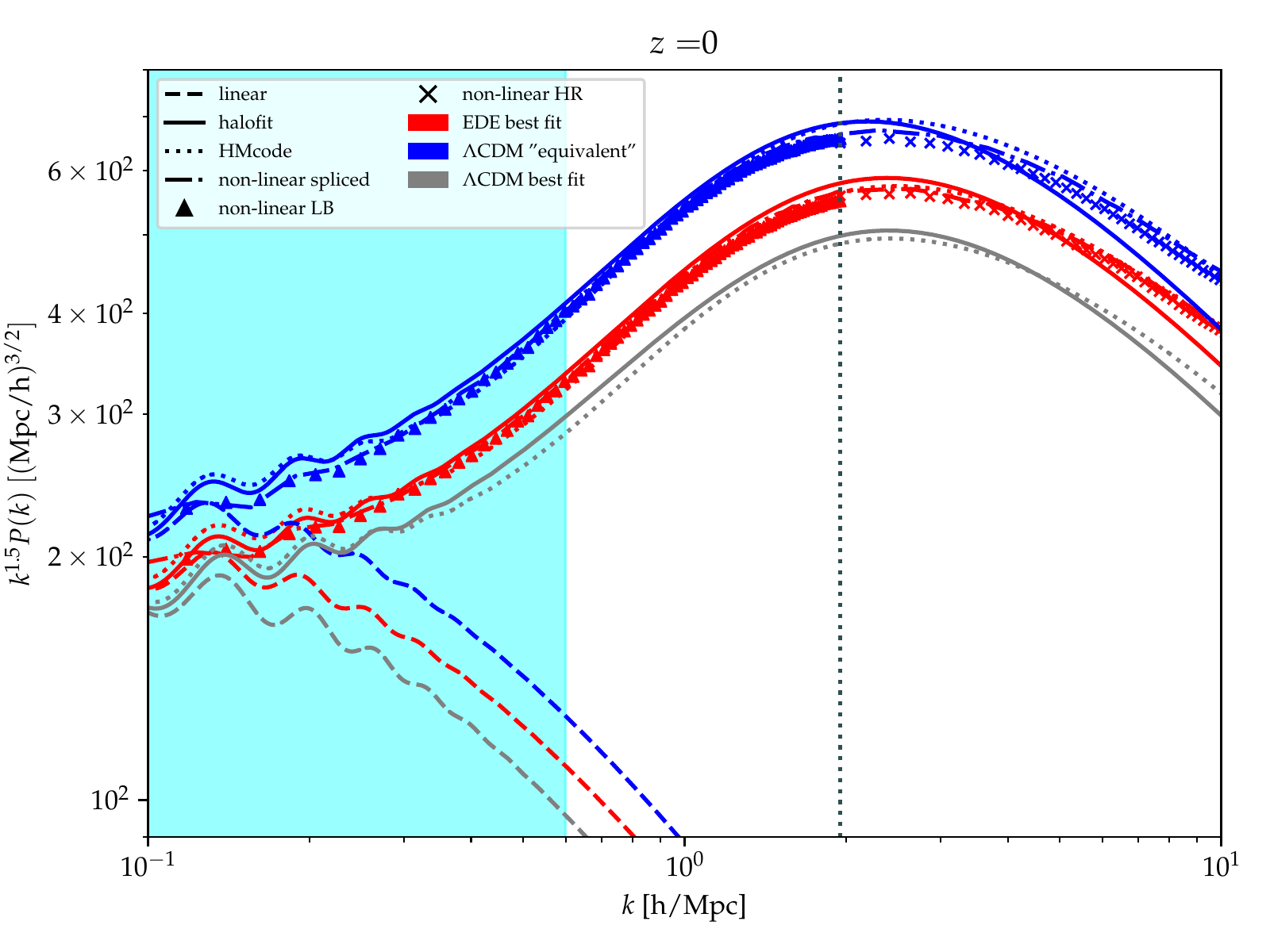} &
\includegraphics[width=.499\textwidth]{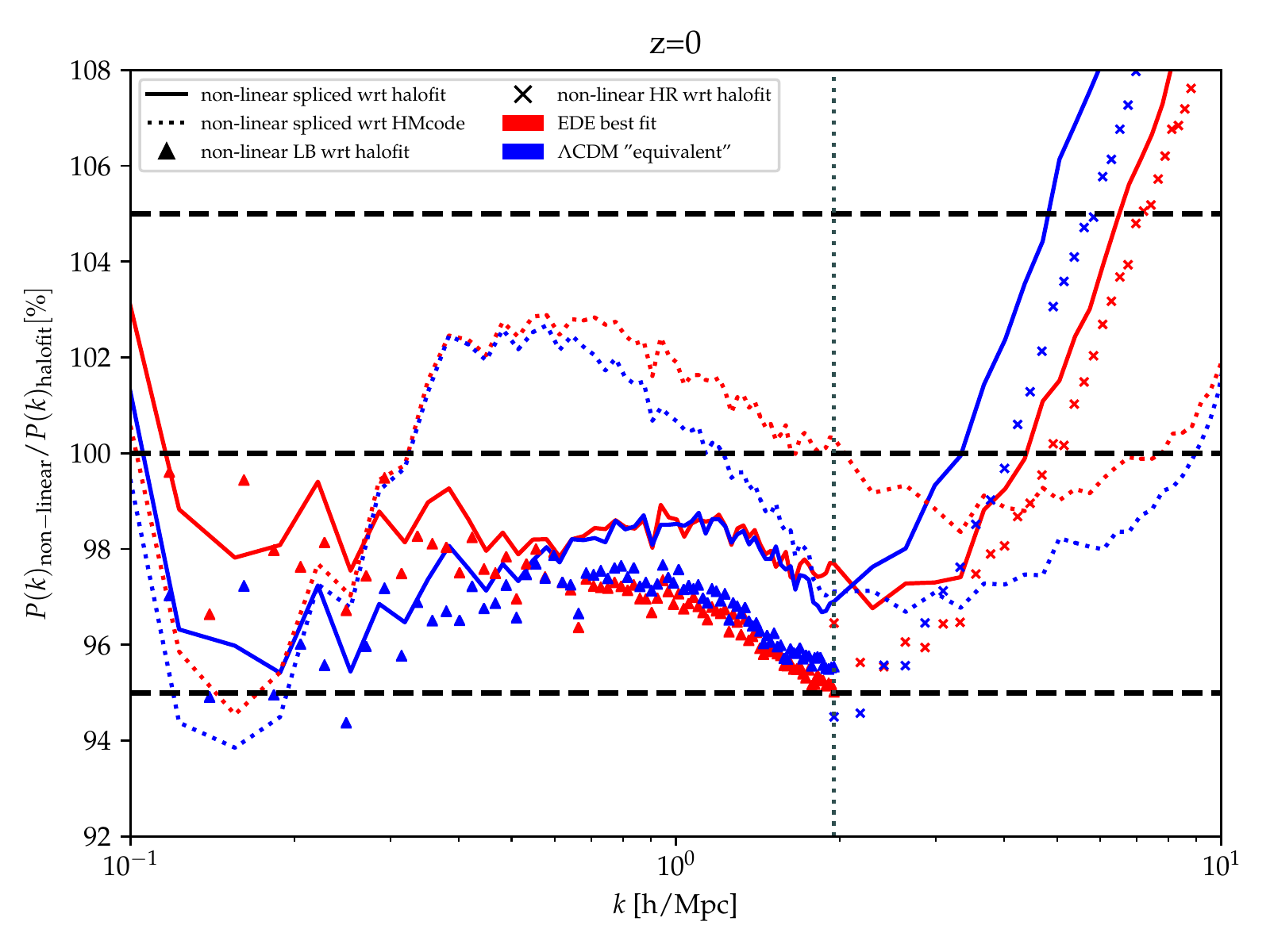}\\
\end{tabular}

\includegraphics[width=.499\textwidth]{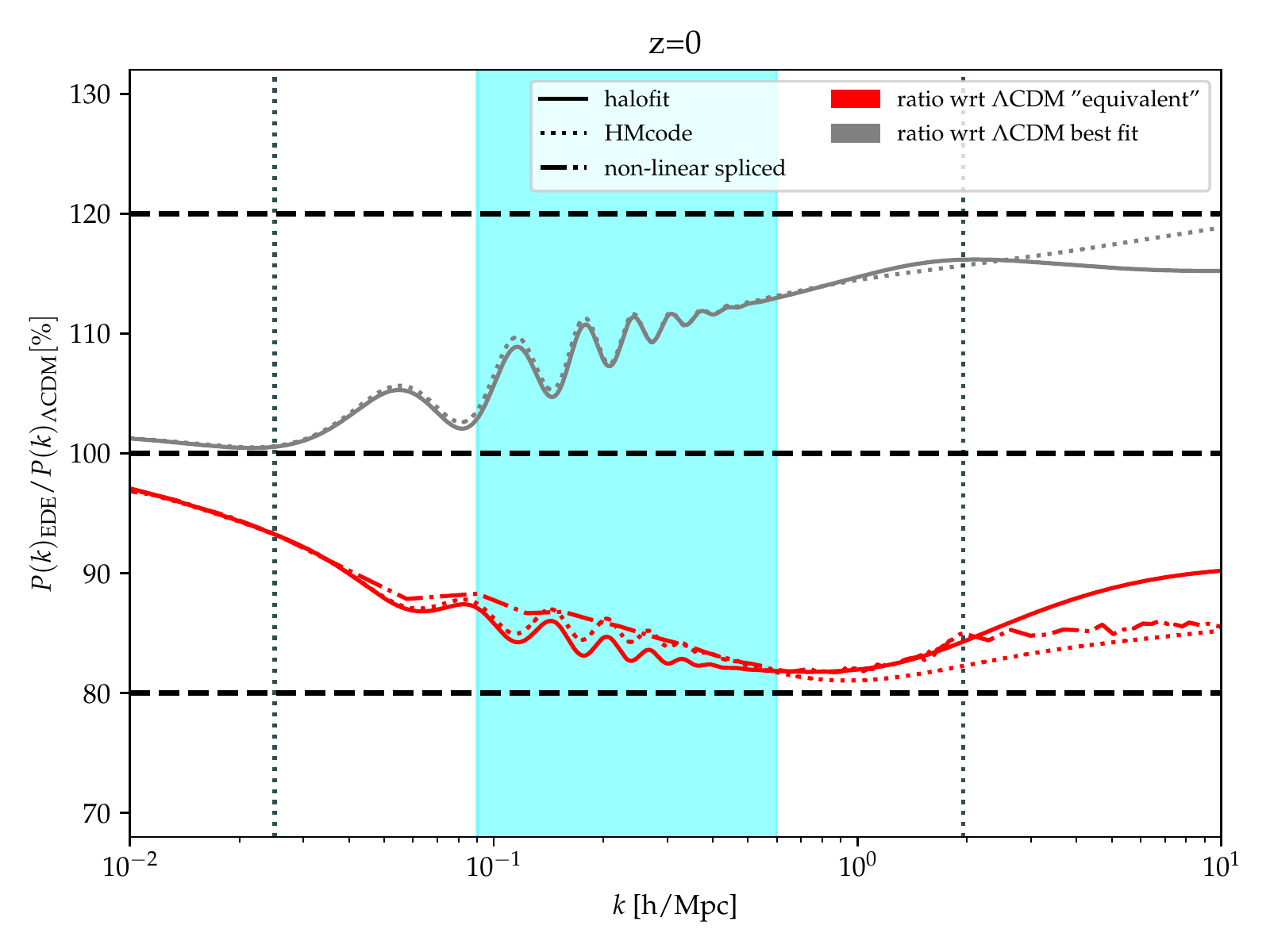}
\caption{In the top left panel we show the matter power spectra
extracted from our simulations, and the ones computed with {\sc{halofit}}/{\sc{HMcode}}. The blue curves refer to the $\Lambda \rm CDM$ scenario, while the red ones refer to the EDE best fit model. We also report the best fit $\Lambda \rm CDM$ case from Planck 2018.
The spliced power spectra are reported as thick dot-dashed lines. Symbols stand for the outputs of the LB and HR simulations.
The solid/dotted lines are the non-linear power spectra from {\sc{halofit}}/{\sc{HMcode}}, whereas the dashed lines are the corresponding linear power spectra used to produce the initial conditions for the simulations. The cyan shaded band roughly corresponds to the scales probed by DES-Y1. 
In the top right panel, we show the ratio between the non-linear matter power spectra from our simulations and the ones computed with {\sc{halofit}}/{\sc{HMcode}}, for both the \LCDM~``equivalent'' and the EDE best fit models, adopting the same linestyle-code and color-code.
In the bottom right panel we compare departures from the \LCDM~model in terms of ratios of non-linear matter power spectra, adopting the same linestyle-code and color-code. 
}
\label{fig:sims1}
\end{figure*}


The goal of this Section is to show that the impact on the non-linear matter power spectrum do to the presence of EDE is mainly due to changes in the standard $\Lambda$CDM free parameters with respect to their reference values, and therefore such impact can be safely studied without further modifying or re-calibrating {\sc{halofit}}/{\sc{HMcode}}.
\begin{table*}[!t]
    \centering
    \begin{tabular}{|l|c|c|c|c|}
        \hline \hline
         Model & Particles ($N$) & Box size ($L$) & Mass resolution & Label \\ \hline \hline
         $\Lambda$CDM/EDE & $1024^3$ & 250 $h^{-1}$ Mpc & $ 1.2 \cdot 10^{9} \, h^{-1} \, \rm M_\odot$ & HR  \\   
         $\Lambda$CDM/EDE & $1024^3$ & 1000 $h^{-1}$ Mpc & 7.5 $\cdot 10^{10} \, h^{-1} \, \rm M_\odot $ & LB  \\
         $\Lambda$CDM/EDE & $256^3$  & 250 $h^{-1}$ Mpc & 7.5 $\cdot 10^{10} \, h^{-1} \, M_\odot $ & LR  \\ \hline  
    \end{tabular}
    \caption{Summary of the properties of the cosmological simulations used in this work. Notice that the Figures shown in this Section have been obtained by splicing together (for each redshift and model) the non-linear matter power spectra extracted from the first two simulations listed here, by using the third one to correct for finite-volume and resolution effects (see Appendix~\ref{ap:sims} for details). The labels listed in the last column stand for High Resolution, Large Box, and Low Resolution, respectively.}
    \label{tab:sims}
\end{table*}
To this end, we perform two sets of $N$-Body DM-only simulations (one set for the EDE and one for the $\Lambda$CDM scenarios), as reported in Table \ref{tab:sims}, by using the $N$-body code {\sc{GADGET-3}}, a modified version of the publicly available numerical code {\sc{GADGET-2}}~\cite{Springel:2000qu,Springel:2005mi}. The initial conditions have been produced by displacing the DM particles from a cubic Cartesian grid according to second-order Lagrangian Perturbation Theory, with the {\sc{2LPTic}} public code~\cite{Crocce:2006ve}, at redshift $z=99$. The corresponding input linear matter power spectra, for both the EDE and $\Lambda$CDM cases, were computed with {\sc {AxiClass}}, the aforementioned modified version \cite{Smith:2019ihp} of the publicly available code {\sc {CLASS}} \cite{Blas:2011rf}. For all of the simulations, we kept the cosmological parameters fixed to their EDE best fit values from Ref.~\cite{Smith:2019ihp} (very close to ours), namely $H_0 = 72.81$, $\Omega_m = 0.2915$, $A_s = 2.191 \cdot 10^{-9}$, $n_s = 0.986$ for both cosmological scenarios; plus the additional parameters $\log_{10}(z_c) = 1.04106$, $f_{\rm EDE}(z_c) = 0.132$, $\Theta_i = 2.72$, $n=2.6$ for the EDE model.
To bind together the matter power spectra extracted from simulations with different resolutions we adopt a splicing technique described in detail in Appendix~\ref{ap:sims}. Our results are summarized in Fig.~\ref{fig:sims1}, and we refer to Appendix~\ref{ap:sims} for a deeper technical discussion.

In the top panel of Fig.~\ref{fig:sims1} we compare the matter power spectra extracted from our simulations, with the ones computed with {\sc{halofit}}/{\sc{HMcode}}, at redshift $z=0$.
The blue curves refer to the $\Lambda \rm CDM$ scenario -- dubbed hereafter as $\Lambda$CDM ``equivalent'' --  while the red ones refer to the EDE best fit model. As a reference, we also report the best fit $\Lambda \rm CDM$ case from Planck 2018. The spliced power spectra are denoted by thick dot-dashed lines. Symbols stand for the output power spectra of the ``non-spliced'' LB and HR simulations.
The solid/dotted lines are the non-linear power spectra from {\sc{halofit}}/{\sc{HMcode}}, while the dashed lines are the corresponding linear power spectra used to set the initial conditions for the simulations. In the right panel, we adopt the same linestyle-code and color-code to show the ratio between the non-linear power spectra produced by {\sc{halofit}}/{\sc{HMcode}} with respect to the ones extracted from our simulations.
The thick horizontal lines highlight $\pm 5\%$ deviations.
In Appendix~\ref{ap:sims} we extend the analysis to three additional redshift bins -- $z = 0.5, 1.5, 2$ -- obtaining analogous results. We can thus conclude that the differences between {\sc{halofit}}/{\sc{HMcode}} predictions with respect to the outputs of our $N$-Body simulations are below $5\%$ level, for scales 
$10^{-2} \lesssim  k \lesssim 10$~$h/$Mpc, at redshifts $ 0 \leq z \leq 2$, for both $\Lambda \rm CDM$ and EDE models. Whereas this is a very well established result for the $\Lambda \rm CDM$ paradigm, this is not often the case for alternative cosmological scenarios, such as the one considered in this work.
Let us note that the exponential increase in the difference between the outputs by simulations and {\sc{halofit}} at $k \sim\, 10~h/$Mpc is absent when one compares the outputs from simulations with the predictions by {\sc{HMcode}}. As expected, the latter method is more accurate than {\sc{halofit}} in modelling the very small-scale and very low-$z$ regime.
In this work, we therefore make use of {\sc{HMcode}} to model the non-linear evolution of perturbations, following the approach adopted by the KiDS collaboration.

We also present our results in terms of ratios between the matter power spectra in the EDE and the $\Lambda \rm CDM$ models in Fig.~\ref{fig:sims1} bottom panel. The comparison between the EDE best fit and the $\Lambda$CDM equivalent confirms
that it is not the intrinsic presence of EDE that enhances the matter power spectrum on small scales, exacerbating the $S_8$ tension. Rather, the EDE reduces the growth of perturbations at fixed $\omega_{\rm cdm}$. As already pointed out, such power enhancement is instead due to variations in the standard $\Lambda$CDM parameters -- mostly an increase in $\omega_{\rm cdm}$ -- induced to balance the EDE impact on the CMB. This suggests that the limitations of the EDE are not intrinsic to its presence, but rather to an accidental degeneracy that could be alleviated in an extended model. This will be the starting point of Section~\ref{sec:ext}, where we will outline possible paths towards restoring the agreement with WL measurements in (extended) EDE cosmologies. 

In view of these considerations, it is straightforward to conclude that LSS surveys constitute an ideal counterpart to CMB data, given the complementarity between the regimes that they probe.
However, in section \ref{subsec:mcmc} we will show that currently available weak lensing data are not sensitive enough to unequivocally capture the signature of EDE.
This will clearly not be the case when more precise data (e.g. from Euclid~\cite{Amendola:2016saw}) will become available.
As also our results suggest, it will soon be necessary to go beyond the {\sc{halofit}}/{\sc{HMcode}} prescription for modelling the non-linear power spectrum (see e.g.~\cite{Chudaykin:2020aoj,Ivanov:2020ril,DAmico:2020ods,Klypin:2020tud}).
Furthermore, it might be already possible to test $\mathcal{O}(20\%)$ deviations in the small-scale power, as the ones shown in the bottom panel of Figure~\ref{fig:sims1}, with current Lyman-$\alpha$ forest flux power spectrum data \cite{Murgia:2017lwo,Murgia:2018now,Murgia:2019duy,Archidiacono:2019wdp,Miller:2019pss,Baldes:2020nuv} and the EFT of LSS data analysis of BOSS data (see e.g. Refs.~\cite{Ivanov:2020ril,DAmico:2020ods} for a recent analysis in the 3-parameter model). We leave these tasks for future work.

\subsection{MCMC analysis against weak lensing data}
\label{subsec:mcmc}

\begin{figure*}
    \centering
\includegraphics[scale=0.7]{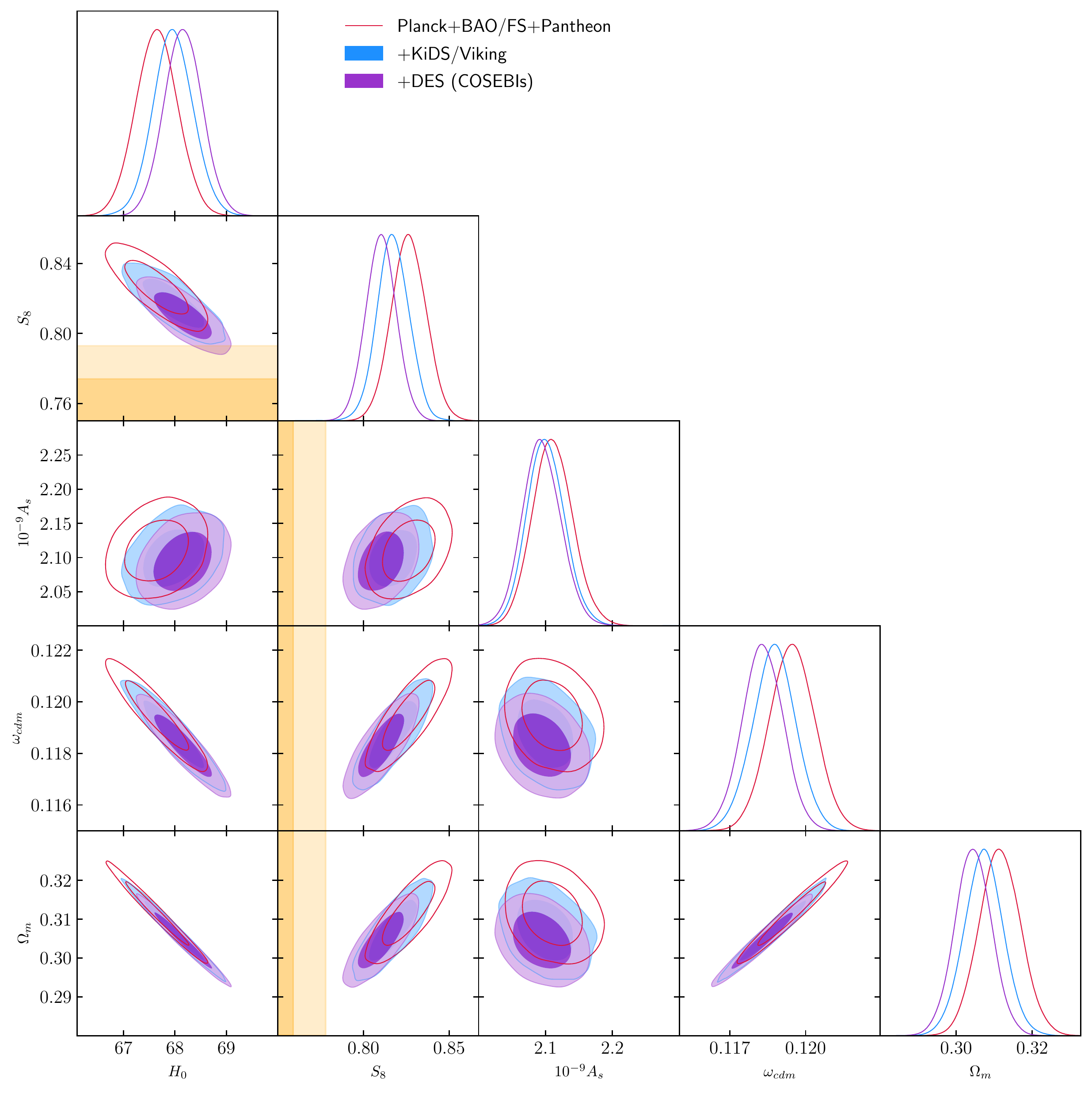}
    \caption{Reconstructed 2D posterior distributions of a subset of parameters for various data set combinations (see legend) in the $\Lambda$CDM cosmology. }
    \label{fig:posterior-lcdm}
\end{figure*}

\begin{figure*}
    \centering
\includegraphics[scale=0.7]{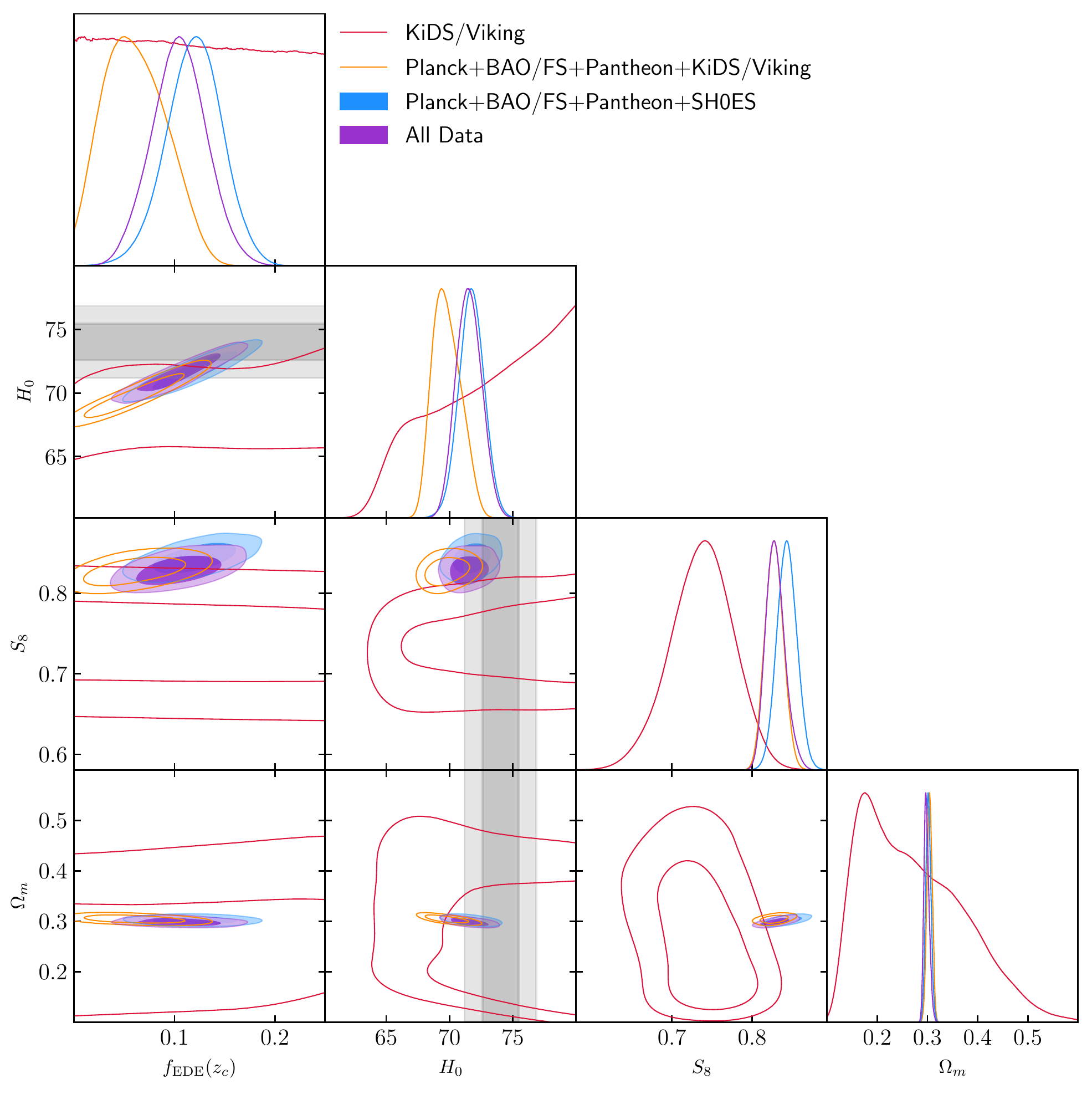}
    \caption{Reconstructed 2D posterior distributions of a subset of parameters for various data set combinations (see legend) in the 1-parameter EDE cosmology. }
    \label{fig:posterior-1param-kids}
\end{figure*}

\begin{table*}
\scalebox{0.9}{
  \begin{tabular}{|l|c|c|c|c|}
          \hline Model & \multicolumn{2}{c|}{$\Lambda$CDM} & \multicolumn{2}{c|}{EDE}\\
          \hline Parameter &Base+{\sc KiDS/Viking}&+\SHOES&Base+{\sc KiDS/Viking}&+\SHOES\\ \hline \hline
    $H_0$ [km/s/Mpc] &  $67.97\pm0.38$ & $68.4\pm0.38$ & $69.75(68.95)_{-1.1}^{+0.99}$   & $71.58(72.22)_{-0.97}^{+1}$  \\
    $100~\omega_b$  & $2.248(2.248)\pm0.013$ & $2.257(2.256)\pm0.013$& $2.261(2.253)_{-0.015}^{+0.014}$& $2.277(2.282)_{-0.015}^{+0.013}$  \\
   $\omega_{\rm cdm}$ & $0.1187(0.1188)\pm0.0009$ & $0.1179(0.1180)\pm0.0009$& $0.1245(0.1235)_{-0.0039}^{+0.0028}$& $0.1291(0.1310) \pm 0.0034$  \\
    $10^9A_s$  &  $2.10(2.08)\pm0.028$ & $2.11(2.12)_{-0.032}^{+0.03}$  &$2.117(2.116)_{-0.033}^{+0.03}$ &$2.136(2.130)_{-0.033}^{+0.03}$  \\
    $n_s$ & $0.9685(0.9667)_{-0.0036}^{+0.0038}$ & $0.9708(0.9691)_{-0.0035}^{+0.0039}$ &$0.9778(0.9740)_{-0.0068}^{+0.0061}$  &  $0.9872(0.9907)_{-0.0055}^{+0.0064}$ \\
    $\tau_{\rm reio}$  & $0.0556(0.0520)_{-0.0066}^{+0.0069}$  & $0.0585(0.0589)_{-0.0075}^{+0.0074}$&$0.0547(0.0559)_{-0.0074}^{+0.0067}$  & $0.0552(0.0536)_{-0.0073}^{+0.0068}$\\
    $f_{\rm EDE}(z_c)$ & $-$ & $-$ &  $0.058(0.042)_{-0.034}^{+0.028}$ & $0.104(0.122)_{-0.025}^{+0.029}$\\
    \hline
    $100~\theta_s$  &  $1.04198(1.04165)\pm0.00028$ &  $1.04207(1.04210)\pm0.00028$&$1.04165(1.04181)_{-0.00032}^{+0.00035}$ & $1.04146(1.04130)_{-0.00034}^{+0.00031}$  \\
     $S_8$ & $0.8172(0.8137)_{-0.0096}^{+0.009}$ & $0.8092(0.8094)_{-0.0098}^{+0.0091}$ &$0.826(0.831) \pm 0.011$ & $0.829(0.828)_{-0.011}^{+0.012}$\\
    $\Omega_m$ & $0.307(0.309)\pm0.005$&$0.302(0.302)\pm0.005$ &$0.3037(0.3085)_{-0.0055}^{+0.0054}$ & $0.2976(0.2962)_{-0.0051}^{+0.005}$ \\
    \hline
    $\chi^2_{\rm min}$ & 3996.82&4011.16& 3992.11 & 3997.67 \\
    \hline
  \end{tabular}
  }
  \caption{The mean (best-fit) $\pm1\sigma$ error of the cosmological parameters reconstructed from the combined analysis of KIDS/Viking with other data. The `Base' dataset refers to \Planck+\BAO/\FS+\Pantheon.  We also report the $\chi^2_{\rm min}$ for each model and data set combination.}
  \label{table:kids}
\end{table*}

In the following, we will focus on the 1-parameter EDE cosmology, fixing $\Theta_i$ and \Logzc~to their best fit values from \Planck~only. Firstly, we test the model against the \KV~cosmic shear measurements. We follow the prescription described in Ref.~\cite{Hildebrandt:2018yau} and make use of the {\sc HMcode} algorithm~\cite{Mead:2015yca} (with 9 nuisance parameters) to model the non-linear matter power spectrum. Secondly, we perform an analysis trading \KV~data for a split-normal likelihood on $S_8$ as inferred from the joint \KV+{\sc DES} data using Complete Orthogonal Sets of E/B-Integrals (COSEBIs), namely\footnote{ We stress that $S_8$ is a model-dependent quantity, and it is particularly sensitive to the treatment of the neutrino mass. We therefore make use of the value that was derived following our convention, i.e. at fixed $\sum m_\nu=0.06$ eV.} $S_8 = 0.755^{+0.019}_{-0.021}$ \cite{Asgari:2019fkq}. 
We report results of MCMC analysis of \LCDM~and EDE against the \KV~data and the joint \KV+{\sc DES} data in table \ref{table:kids} and \ref{table:cosebi}.

{\bf Results for \LCDM:} Starting with the $\Lambda$CDM cosmology, we find that combining \Planck~with \KV~data leads to a mild degradation of the $\chi^2_{\rm min}$ of the combined fit: While one might naively expect that the $\chi^2_{\rm min}$ of the global fit should be roughly the sum of the $\chi^2_{\rm min}$ of individual fits, we find that the global $\chi^2_{\rm min}$ is degraded by $\sim+6.5$. Similarly, the inclusion of a tight Gaussian likelihood on $S_8$ as measured by \KV+{\sc DES} leads to a  degradation in the combined $\chi^2\sim+15.5$, while one expects $\sim+1$ for a good fit.  In Fig.~\ref{fig:posterior-lcdm}, we show the reconstructed 2D posteriors of $\{H_0,S_8,10^{-9}A_s,\omega_{\rm cdm}, \Omega_m\}$ in the \LCDM~model. One can see that the degradation in $\chi^2_{\rm min}$ is accompanied by shifts in the mean of any parameter correlated with $S_8$, in particular $A_s$, $\omega_{\rm cdm}$ and $H_0$, without succeeding in getting a good fit to the WL data.
We therefore stress that any of the combined results should be taken with a grain of salt, even in the $\Lambda$CDM framework. This joint analysis serves mostly to demonstrate that the EDE cosmology does not sensibly degrade the fit to the $S_8$ measurement as compared to $\Lambda$CDM, and that currently available WL measurements do not strongly constrain the EDE resolution to the Hubble tension.

{\bf Results for EDE against Planck+KiDS-VIKING:} In Fig.~\ref{fig:posterior-1param-kids} we show the reconstructed 2D posteriors of $\{f_{\rm EDE}(z_c), H_0,S_8,\Omega_m\}$ in the 1-parameter EDE realization for various data combinations. We start by performing an analysis of EDE against \KV~data only; as expected  we find that the \KV~data have no constraining power on the fraction of EDE.  However, the reconstructed $S_8=0.738^{+0.041}_{-0.038}$ is $\sim$ 2.4$\sigma$ discrepant with that obtained from previous analyses, suggesting a potential discordance between the cosmologies. For comparison, the prediction for $S_8$ in the  $\Lambda$CDM model obtained from \Planck~data is $2.3\sigma$ discrepant with that from \KV~data \cite{Hildebrandt:2018yau}. Therefore, although the mean value has increased, the level of the $S_8$ tension in the EDE cosmology is similar to that in $\Lambda$CDM because of larger error bars. Combining \KV~to~\Planck+\BAO+\Pantheon+\FS, a non-zero EDE contribution is still favored at $\sim~1.5\sigma$, but the reconstructed mean fraction has moved downwards by $\sim 0.7\sigma$. This was expected, given the positive correlation between \fEDE~and $S_8$. In this cosmology, \Planck~data are still slightly better fitted ($\Delta\chi^2_{\rm min}\sim-6$) than in \LCDM, while the fit to \KV~data is degraded by $\sim+2$. Once a prior on $H_0$ from \SHOES~is added, we find again $f_{\rm EDE}(z_c)\sim10\pm3\%$, at the cost of increasing the total $\chi^2_{\rm min}\sim+5.5$. The increase in $\chi^2$ is partly due to the inclusion of \SHOES~($\chi^2 \sim 1.62$, a reasonably good fit), and also to a mild degradation in the fit to \Planck~($\sim+3$) and BAO ($\sim+1.6$). The reason is that the inclusion of \KV~data reduces the degeneracy between \fEDE~and the $\Lambda$CDM parameters, in particular the one with $\omega_{\rm cdm}$. Note that the goodness of \Planck~fit is not sensibly degraded as compared to \LCDM, since the $\chi^2$ stays better than that from \LCDM~fitted on \Planck~only.
In fact, when compared to $\Lambda$CDM, the combined $\chi^2$ is improved by $\sim-13$ (for one extra parameter), indicating a significant preference for EDE despite the presence of \KV~data. Looking at the individual $\chi^2_{\rm min}$, we find indeed that the quality of the fit to \KV~data in the EDE cosmology that resolves the Hubble tension is barely changed ($\Delta\chi^2\sim+1.6$ for 195 data points \cite{Hildebrandt:2018yau}) compared to the $\Lambda$CDM fit to the same data set.

\begin{figure*}
    \centering
\includegraphics[scale=0.7]{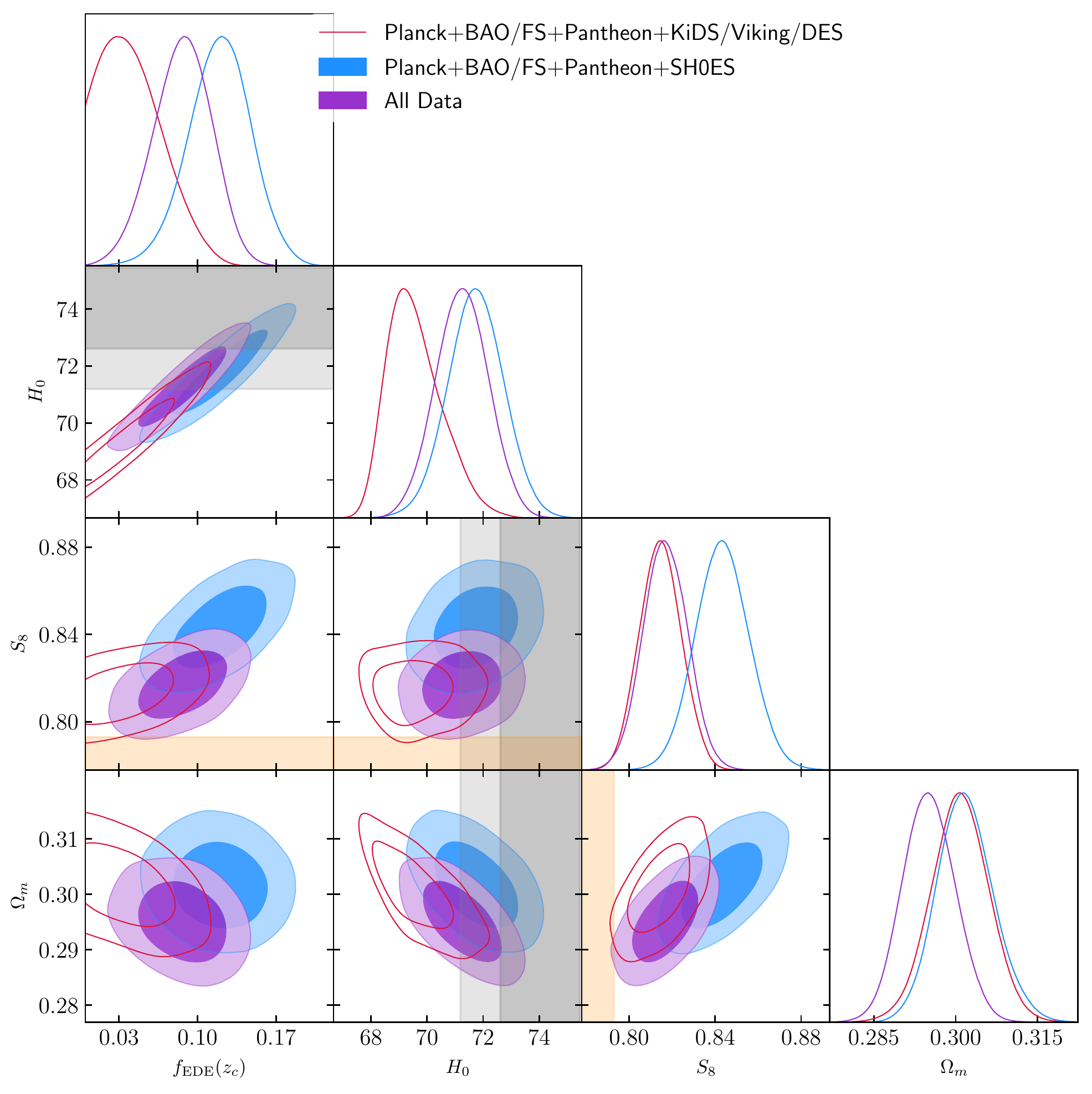}
    \caption{Reconstructed 2D posterior distributions of a subset of parameters for various data set combinations (see legend) in the 1-parameter EDE cosmology.}
    \label{fig:posterior-1param-cosebi}
\end{figure*}

\begin{table*}[t!]
\scalebox{0.9}{
  \begin{tabular}{|l|c|c|c|c|}
     \hline
     Model & \multicolumn{2}{c|}{\LCDM} & \multicolumn{2}{c|}{EDE} \\
     \hline Parameter &Base+{\sc KiDS/Viking/DES}&+\SHOES&Base+{\sc KiDS/Viking/DES}&+\SHOES\\ \hline \hline
    $H_0$ [km/s/Mpc] &$68.16(68.15) \pm 0.38$ &$68.56(68.69)_{-0.39}^{+0.38}$  & $69.56(69.55)_{-1.2}^{+0.72}$ & $71.29(71.81)_{-0.94}^{+0.94}$ \\
    $100~\omega_b$  & $2.251(2.253) \pm 0.013$ & $2.26(2.263)_{-0.014}^{+0.013}$ &$2.262(2.270)_{-0.015}^{+0.014}$  &  $2.278(2.288) \pm 0.014$\\
    $\omega_{\rm cdm}$& $0.1183(0.1183)_{-0.00082}^{+0.00084}$ & $0.1175(0.1172)_{-0.00083}^{+0.00085}$ &$0.1223(0.1199)_{-0.0036}^{+0.002}$ & $0.1264(0.1270)_{-0.0032}^{+0.003}$  \\
    $10^9A_s$ &  $2.094(2.091)_{-0.03}^{+0.029}$ & $2.104(2.115)_{-0.032}^{+0.029}$&$3.046(2.107)_{-0.015}^{+0.014}$&$2.121(2.117) \pm 0.031$\\
    $n_s$&  $0.9691(0.9705) \pm 0.0037$ &  $0.9712(0.9731) \pm 0.0037$ &$0.9765(0.9782)_{-0.0065}^{+0.0051}$ &  $0.9854(0.9892)_{-0.0057}^{+0.0055}$ \\
    $\tau_{\rm reio}$ &$0.0546(0.0538)_{-0.0073}^{+0.0069}$ & $0.0576(0.0602)_{-0.0077}^{+0.0069}$&$0.05339(0.0559)_{-0.0072}^{+0.0071}$&  $0.05441(0.05254)_{-0.0072}^{+0.007}$ \\
    $f_{\rm EDE}(z_c)$  & $-$ &$-$ &$<0.094 (0.029)$& $0.087(0.097)_{-0.024}^{+0.029}$\\
    \hline
    $100~\theta_s$ &  $1.04198(1.04195)_{-0.00029}^{+0.00028}$ & $1.04207(1.04209)_{-0.00029}^{+0.00028}$&$1.04178(1.04190)_{-0.00031}^{+0.00032}$ & $1.04157(1.04149)_{-0.00032}^{+0.00033}$ \\
    $S_8$ &   $0.8043(0.8102)_{-0.0057}^{+0.0055}$ &$0.8039(0.8023)_{-0.0058}^{+0.0056}$  &$0.8145(0.8036)_{-0.01}^{+0.0098}$ & $0.817(0.812)_{-0.011}^{+0.01}$\\
    $\Omega_m$ &  $0.3045(0.3046)_{-0.005}^{+0.0048}$ & $0.2994(0.2978) \pm 0.0049$  &$0.3008( 0.2961)_{-0.0053}^{+0.0054}$ &  $0.2949(0.2919)_{-0.005}^{+0.0047}$\\
    \hline
    $\chi^2_{\rm min}$ &  3821.93&3837.98 & 3820.46 & 3826.35 \\
    \hline
  \end{tabular}
  }
  \caption{The mean (best-fit) $\pm1\sigma$ error of the cosmological parameters reconstructed from the  combined analysis of the KIDS/Viking/DES data with other data discussed in the text, with and without a prior on $H_0$ from \SHOES. The `Base' dataset refers to \Planck+\BAO/\FS+\Pantheon. We also report the $\chi^2_{\rm min}$ for each model and data set combination.}
  \label{table:cosebi}
\end{table*}

{\bf Results for EDE against Planck+KIDS+DES:} We now trade \KV~data for a split-normal likelihood on $S_8$ as inferred from the joint \KV+{\sc DES} data. We note that the tension between the value of this joint $S_8$ measurement and that predicted by our fiducial EDE model (obtained from the global fit of \Planck+\BAO+\FS+\Pantheon+\SHOES) is at the $\sim3.8\sigma$ level (slightly increased from 3.2$\sigma$ tension in $\Lambda$CDM). It would be interesting to  quantify the level of tension between these data sets in the $\Lambda$CDM framework using more robust statistical tools than the `difference in the mean' used here, as done for instance in Refs.~\cite{Raveri:2018wln,Handley:2019wlz}. However, we note that authors from Ref.~\cite{Raveri:2018wln} found that the less precise \KV~data available at that time were already in significant statistical disagreement with the prediction from \LCDM. We anticipate that this more robust approach would strengthen the case for a statistically significant discrepancy, even in $\Lambda$CDM, and therefore the need to apply caution when drawing conclusions from the combined analysis. We show the reconstructed 2D posteriors of $\{f_{\rm EDE}(z_c), H_0,S_8,\Omega_m\}$ in the 1-parameter EDE model in Fig.~\ref{fig:posterior-1param-cosebi}. 
Without the \SHOES~prior, \fEDE~is compatible with 0 at $1\sigma$, and we find an upper limit on $f_{\rm EDE}(z_c)<0.094$ at 95\% C.L. This constraint is significantly weaker than that derived in Refs~\cite{Hill:2020osr}, despite the fact that we have {\em reduced} the parameter space. We have simply adopted a different `prior' choice on $\Theta_i$ and \Logzc~(i.e. here we fix them), demonstrating that the current constraints from WL -- besides being derived from statistically inconsistent data set -- are not robust. Looking at $\chi^2_{\rm min}$, we find that the resulting best fit cosmology degrades the fit to Planck by $\sim+6$ while providing a poor fit to the $S_8$ likelihood ($\chi^2 = 8.3$ for a single data point). Still, the best-fit is marginally better than that of $\Lambda$CDM adjusted on the same sets of data ($\Delta\chi^2_{\rm min}\sim-1.5$). 
Once we include the \SHOES~prior, we find again \fEDE~to be non-zero at more than 3$\sigma$,~\fEDE$\simeq9\pm3\%$,  with a global $\Delta\chi^2_{\rm min} \simeq -11.6$. Looking at individual $\chi^2_{\rm min}$, we find that the fit to \PlanckTTTEEE, \PlanckLensing, \BAO~and \FS~data is somewhat degraded compared to the best fit EDE cosmology obtained without $S_8$ prior, as a consequence of the breaking of the \fEDE$-\omega_{\rm cdm}$ degeneracy.  However, as expected, we note that the $S_8$ likelihood has a $\chi^2\simeq 9$, which is not particularly worst that the one obtained in the $\Lambda$CDM case without \SHOES~($\chi^2\simeq8.3$). This indicates that any constraint on the EDE derived from this combined analysis should be regarded with caution, as the cosmology reconstructed from the analysis does not provide a good fit to the $S_8$ data. This naturally impacts the reconstructed $H_0$, which is $\sim0.6\sigma$ lower than without the $S_8$ likelihood, although the fit to \SHOES~is still reasonably good ($\chi^2 \simeq 2.4$). We therefore conclude that current $S_8$ measurements do not exclude the EDE resolution to the Hubble tension; however, they do call for new physics beyond EDE -- or unknown systematics -- to explain the intriguingly low measured $S_8$ values. 
\subsection{1pEDE vs $\Lambda$CDM: a Bayesian view}

\begin{table*}[!t]
  \begin{tabular}{|l|c|c|c|c|}
  \hline
   & Base&Base+\SHOES&Base+{\sc KiDS/Viking/DES}&Base+{\sc KiDS/Viking/DES}+\SHOES\\
   \hline
 $\Delta{\rm log} B$ & +4 & -0.7 & +5.1 & +2.2 \\ \hline  
\end{tabular}
\caption{Difference in the Bayesian evidence $\Delta{\rm log} B = {\rm log} B({\rm EDE})-{\rm log} B(\Lambda{\rm CDM}$) between the 1pEDE and $\Lambda$CDM cosmology for the data combination as indicated by the row title. The `Base' dataset refers to \Planck+\BAO/\FS+\Pantheon.\label{table:evidence}}
\end{table*}

{ 

In order to get a sense of whether 1pEDE is favored over $\Lambda$CDM, we perform a Bayesian analysis against 4 representative data combination, namely i) Base; ii) Base+\SHOES; iii) Base+{\sc KiDS/Viking/DES}; iv) Base+\SHOES+{\sc KiDS/Viking/DES}. We recall that our `Base' dataset refers to \Planck+\BAO/\FS+\Pantheon. We make use of the sampler {\sc MultiNest} \cite{Feroz:2008xx}, with 800 live points and a tolerance condition on the evidence for stopping the sampling equal to 0.1. We perform model comparison by calculating $\Delta{\rm log} B = {\rm log} B({\rm EDE})-{\rm log} B(\Lambda{\rm CDM})$ as reported in Table~\ref{table:evidence}. As can be anticipated from the small $\Delta\chi^2$ values, we find that the 1pEDE is not favored in a Bayesian sense over $\Lambda$CDM, expect in the case \Planck+\BAO/\FS+\Pantheon+\SHOES, where the preference is `weak' according to Jeffrey's scale. We note that Ref.~\cite{Poulin:2018cxd} reported `strong evidence' in favor of EDE for the same data combination. We checked that the differences can be attributed to (in order of importance): i) the new Planck 2018 data; ii) a slightly less restrictive prior on $f_{\rm EDE}$, taken to vary between $[0,0.5]$ instead of $[0,0.3]$; iii) the correct inclusion of BAO/FS data from BOSS DR12.}

\section{A common resolution to the $H_0$ and $S_8$ tensions in the EDE cosmology?}
\label{sec:towards}
\subsection{EDE and the $S_8$ tension in light of Planck unlensed CMB spectrum}
It has been noted that there exists a number of `curiosities' in \Planck~that can potentially shed light on cosmological tensions. In particular, there is a residual oscillatory feature in the \Planck~TT data at $1100\lesssim\ell\lesssim2000$ compared to the best fit $\Lambda$CDM prediction~\cite{Aghanim:2016sns,Aghanim:2018eyx}.
This feature can be captured by an extra source of smoothing of the acoustic peaks, as modelled by the `$A_{\rm lens}^{\rm \phi\phi}$' parameter which is used to re-scale the amplitude of the lensing potential power spectrum $C_l^{\phi\phi}\to A_{\rm lens}^{\rm \phi\phi}C_l^{\phi\phi,}$ at every point in parameter space. However, the amplitude of the lensing potential power spectrum can also be estimated directly from the lensing-reconstruction and is compatible with the $\Lambda$CDM expectation, such that while this extra smoothing looks like lensing, it cannot be attributed to actual gravitational lensing.

A thorough investigation of the lensing-like tensions in the Planck legacy release was performed in Refs.~\cite{Aghanim:2016sns,Motloch:2018pjy,Motloch:2019gux}. It has been noted in particular that, once marginalizing over the lensing information, the `unlensed' CMB temperature and polarization power spectra favor a cosmology with a lower $A_s$ and $\Omega_{\rm cdm}h^2$. Indeed, these parameters are strongly correlated with the amplitude of the lensing power spectrum, such that the lensing-like anomaly pulls up these parameters. Additionally, since the acoustic feature of the CMB tightly constraints the parameter combination $\Omega_mh^3$, a lower $\Omega_{\rm cdm}h^2$ is compensated by a higher $H_0$. As a consequence, the unlensed $\Lambda$CDM cosmology shows no $S_8$ tension, and a milder (although still $>3.5\sigma$ significant) $H_0$ tension. It was also pointed out that this `unlensed' cosmology is in good agreement with the $\Lambda$CDM cosmology reconstructed from the SPTPol data \cite{Aghanim:2016sns,Henning_2018,Chudaykin:2020acu}.

\begin{figure*}
    \centering
    \includegraphics[scale=0.5]{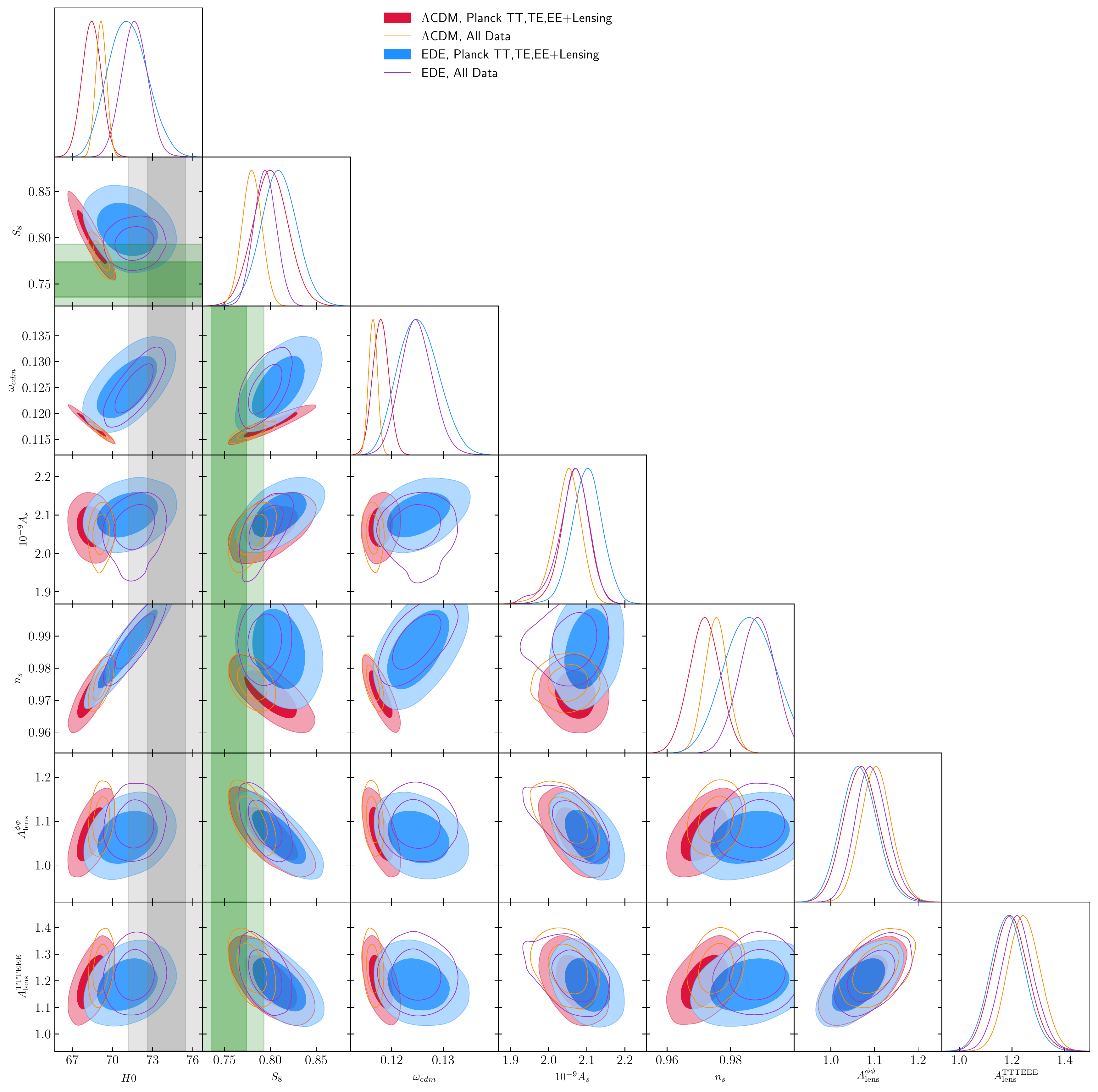}
    \caption{Reconstructed 1- and 2D-posteriors of a subset of parameters in the $\Lambda$CDM and 1-parameter EDE cosmology for various data sets (see legend),  once marginalizing over $A_{\rm lens}$ and $A_{\rm lens}^{\phi\phi}$}
    \label{fig:lcdm_fede_alens}
\end{figure*}

\begin{figure*}
    \centering
    \includegraphics[scale=0.7]{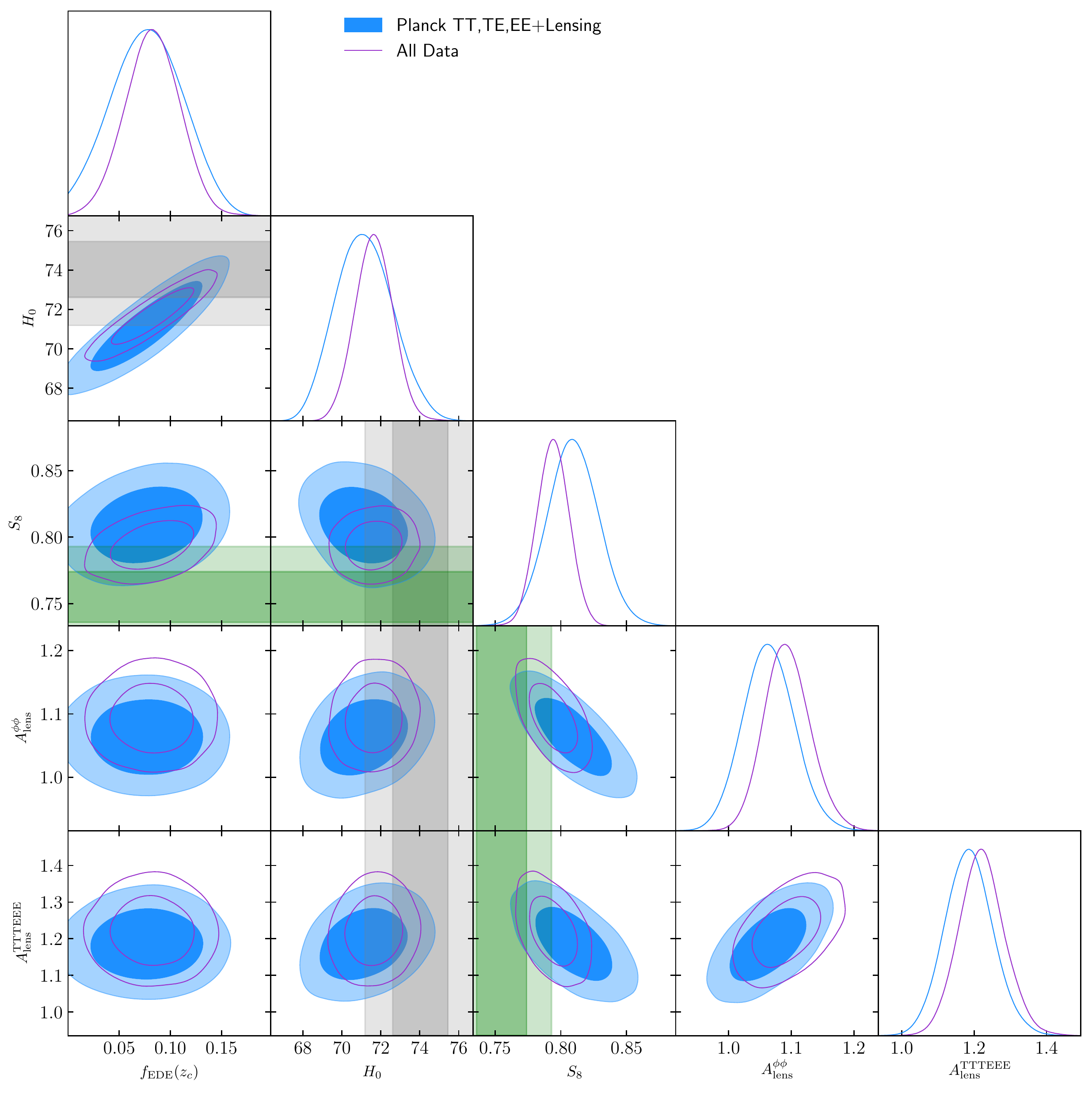}
    \caption{Reconstructed 1- and 2D-posterior of a subset of parameters in the 1-parameter EDE cosmology for various data sets (see legend), once marginalizing over $A_{\rm lens}$ and $A_{\rm lens}^{\phi\phi}$.}
    \label{fig:fede_alens}
\end{figure*}

It is therefore reasonable to ask what is the impact of such anomalies on extensions to $\Lambda$CDM like the EDE under study. To that end, we introduce two additional parameters $A_{\rm lens}$ and $A_{\rm lens}^{\phi\phi}$ whose goal is to marginalize over the lensing information in Planck\footnote{An alternative, more thorough, way is to use CMB lensing principal components as introduced in Ref.~\cite{Motloch:2018pjy,Motloch:2019gux}. As we will show shortly, our reconstructed `unlensed' cosmologies are in good agreement. Our approach follows that introduced in Refs.~\cite{Simard:2017xtw,Wu:2019hek}.}. The latter parameter re-scales  the amplitude of the theory lensing potential power spectrum, while the former only re-scales the amplitude of the acoustic peak smoothing. In practice, the amplitude of the acoustic peak smoothing is then determined by the product $A_{\rm lens}^{\rm TTTEEE}\equiv A_{\rm lens}\times A_{\rm lens}^{\phi\phi}$. We first perform  MCMC analysis of the $\Lambda$CDM and EDE cosmologies against \Planck~data. In a second step, we perform a global analysis combining all the data considered in this work. As before, the joint \KV+{\sc DES} results is modelled via a split-normal likelihood distribution on $S_8$.  The results of these analysis are reported in Tab.~\ref{table:alens} and shown in Figs.~\ref{fig:lcdm_fede_alens} and \ref{fig:fede_alens} .

\begin{table*}[!t]
\scalebox{0.9}{
  \begin{tabular}{|l|c|c|c|c|}
     \hline
     Model & \multicolumn{2}{c|}{\LCDM} & \multicolumn{2}{c|}{EDE} \\
     \hline Parameter &\PlanckTTTEEE+\PlanckLensing&All Data&\PlanckTTTEEE+\PlanckLensing &All Data \\ \hline \hline
    $H_0$ [km/s/Mpc] & $68.44(68.66)_{-0.72}^{+0.74}$&$69.16(69.37)\pm0.41$ &$71.17(72.18)_{-1.6}^{+1.4}$&$71.64(72.07)_{-1}^{+0.94}$  \\
    $100~\omega_b$  & $2.262(2.269)\pm0.018$& $2.277(2.284)\pm0.014$ & $2.284(2.292)\pm0.02$ &$2.292(2.294)\pm0.015$ \\
    $\omega_{\rm cdm}$& $0.1179(0.1174)\pm0.0016$ & $0.1164(0.1159)\pm0.00087$&$0.1253(0.1286)_{-0.0045}^{+0.0037}$ &$0.1248(0.1255)_{-0.0033}^{+0.003}$  \\
    $10^9A_s$ &  $2.069(2.071)_{-0.035}^{+0.038}$  &$2.048(2.053)_{-0.032}^{+0.039}$ & $2.101(2.122)\pm0.041$ &$2.064(2.066)_{-0.033}^{+0.047}$ \\
    $n_s$&  $0.9718(0.9730)\pm0.005$  &$0.9755(0.9786)\pm0.0037$ &$0.9862(0.9925)_{-0.0088}^{+0.008}$ &$0.9884(0.9925)_{-0.0056}^{+0.0059}$\\
    $\tau_{\rm reio}$ & $0.0494(0.0506)_{-0.0079}^{+0.0089}$ & $0.0464(0.0480)_{-0.0074}^{+0.0093}$& $0.0507(0.0529)_{-0.008}^{+0.0087}$  &$0.0429(0.0431)_{-0.0071}^{+0.012}$  \\
    $A_{\rm lens}^{\phi\phi}$ & $1.071(1.075)_{-0.043}^{+0.04}$  &$1.104(1.110)_{-0.038}^{+0.034}$ &   $1.064(1.056)_{-0.043}^{+0.04}$ & $1.093(1.099)_{-0.039}^{+0.035}$ \\
    $A_{\rm lens}^{\rm TTTEEE}$ & $1.195(1.208)_{-0.07}^{+0.066}$ & $1.247(1.266)_{-0.066}^{+0.06}$ & $1.187(1.188)_{-0.07}^{+0.065}$ & $1.222(1.238)_{-0.067}^{+0.061}$  \\
    $f_{\rm EDE}(z_c)$  & $-$ &$-$ & $0.078(0.108)_{-0.038}^{+0.035}$  &$0.082(0.092)\pm0.027$ \\
    \hline
    $100~\theta_s$ &  $1.04205(1.04207) \pm 0.00031$&$1.04215(1.04214)\pm0.00029$ & $1.04165(1.04343)_{-0.00035}^{+0.00036}$ & $1.04165(1.04164)\pm0.00034$ \\
    $S_8$ &  $0.800(0.795)_{-0.02}^{+0.019}$ & $0.780(0.776) \pm 0.011$& $0.801(0.812) \pm 0.02$ & $0.794(0.793)\pm0.013$\\
    $\Omega_m$ &  $0.302(0.297)_{-0.01}^{+0.009}$ &$0.2924(0.2883)_{-0.0051}^{+0.0049}$ & $0.2938(0.2870)_{-0.01}^{+0.0095}$ &$0.2891(0.2870)\pm0.0052$ \\
    \hline
    $\chi^2_{\rm min}$ ($\Lambda$CDM) & 2765.98 & 3816.23 & 2761.98&  3808.40\\
    \hline
  \end{tabular}
  }
  \caption{The mean (best-fit) $\pm1\sigma$ error of the cosmological parameters reconstructed from the lensing-marginalized \Planck~data only and in combination with \BAO/\FS+\Pantheon+\KV-\DES. We also report the $\chi^2_{\rm min}$ for each model and data set combination.}
  \label{table:alens}
\end{table*}

\textbf{Results for $\Lambda$CDM:} We start by analyzing the $\Lambda$CDM cosmology in light of the `unlensed' Planck spectra.  We confirm the results of Refs.~\cite{Aghanim:2018eyx,Motloch:2019gux}: we find that the amount of lensing determined from the peak smoothing $A_{\rm lens}^{\rm TTTEEE}$ is $\sim 2.8\sigma$ higher than the expectation from the $\Lambda$CDM model deduced from the `unlensed' CMB power spectrum. Moreover, the difference between the reconstructed $A_{\rm lens}^{\phi\phi}\simeq1.07\pm0.04$ and $A_{\rm lens}^{\rm TTTEEE}\simeq1.2\pm0.07$ illustrates the fact that this extra smoothing component cannot be due to actual gravitational lensing. Still, this $\Lambda$CDM `unlensed' cosmology is now in much better agreement with the $S_8$ measurements from KiDS and DES, as can be seen in Fig.~\ref{fig:lcdm_fede_alens}. This is due to the fact that the reconstructed $A_s$ and $\omega_{\rm cdm}$ are lower than in the analysis including lensing information. We then perform a global analysis, including all data sets considered in this work. We find that the `unlensed' $\Lambda$CDM cosmology can indeed accommodate a low $S_8$, however this is at the cost of worsening somewhat the fit to \BAO+\FS~ data ($\Delta\chi^2\simeq+3.5$), when compared to the `concordance' $\Lambda$CDM model obtained from a fit to the full Planck data, \BAO~and \FS~(without $S_8$ priors). Additionally, we 
note that accommodating such a low $S_8$ requires a somewhat smaller $\omega_{\rm cdm}$ and $A_s$ (by a little less than $1\sigma$), which are compensated for by pulling up the $A_{\rm lens}^{\phi\phi}$ and $A_{\rm lens}^{\rm TTTEEE}$  by a similar amount. The fit to \SHOES~on the other hand is still very poor, $\chi^2_{\rm min}\simeq 10$, suggesting that the global unlensed cosmology is still in strong tension with \SHOES.

\textbf{Results for EDE:} Turning now to the 1-parameter EDE model, we wish to check whether the EDE cosmology deduced from `unlensed' \Planck~spectra only is in better agreement with both $S_8$  and $H_0$ direct measurements\footnote{A similar study was performed in Ref.~\cite{Motloch:2019gux} for $N_{\rm eff}$. There, it was found that polarization and BAO data exclude $N_{\rm eff}$ as a resolution to the Hubble tension, even after marginalizing over the lensing anomaly in \Planck.}. As one can see from Fig.~\ref{fig:fede_alens}, the lensing-marginalized CMB data does favor non-zero \fEDE~at $\sim2\sigma$ ($\Delta\chi^2_{\rm min}\simeq-4$ with respect to $\Lambda$CDM) and predicts $H_0 \simeq 71.2\pm1.5$ and $S_8\simeq0.81\pm0.02$. Compared to the EDE cosmology reconstructed from the full Planck data, the `tension' with $H_0$ and $S_8$ has therefore decreased  by $\sim 1\sigma$ due to a shift in the mean of the reconstructed posterior in the unlensed cosmology. It is now in $1.4\sigma$ agreement with \SHOES~but stays in mild ($\sim 2.7\sigma$) tension with the combined $S_8$ measurement. The $S_8$ prediction is however in very good agreement with the \KV~and \DES~measurements when considered individually  (an important note since the combined low  $S_8$ value relies on a re-calibration of \DES`photo-metric redshift by the \KV~team). 
Additionally, the $A_{\rm lens}^{\phi\phi}$ and $A_{\rm lens}^{\rm TTTEEE}$ parameters are unchanged with respect to that reconstructed in the $\Lambda$CDM cosmology. Therefore, while the anomalous amount of lensing in \Planck~data is not an effect due to the presence of the EDE, these parameters do not correlate with a non-zero \fEDE, i.e., they do not take values different from the $\Lambda$CDM ones to `hide' the effect of the EDE.

Once all data sets are included in the analysis, a non-zero fraction of EDE is favored at $\sim3.5\sigma$. Interestingly, most of the reconstructed parameters do not shift by more than $\sim 0.5\sigma$; rather, the uncertainty on the reconstructed parameters tighten significantly, as one would expect from making use of additional data. However, similarly to what happens in the $\Lambda$CDM cosmology, the inclusion of the tight-and-low $S_8$ value does force a slightly ($\sim1\sigma$) smaller $A_s$, that is compensated by slightly higher $A_{\rm lens}^{\phi\phi}$ and $A_{\rm lens}^{\rm TTTEEE}$ parameters. The fit to \SHOES~is good ($\chi^2\sim1.9$) and stable when compared to that obtained including Planck lensing information. On the other hand, as expected, the fit to the joint \KV+\DES~$S_8$ is better than in the `lensed' cosmology ($\Delta\chi^2\sim-4.3$), but its value is still somewhat poor ($\chi^2\sim 4$). We emphasize again that the fit to individual $S_8$ measurements, on the other hand, is excellent. If future $S_8$ measurements stay low while becoming more precise, they will be in tension even with the `unlensed' cosmology (whether $\Lambda$CDM or EDE), confirming the need for new physics beyond EDE (or an unknown systematic effect).

\subsection{Extended cosmologies that could help resolving the $S_8$ tension}\label{sec:ext}
Given that both the $\Lambda$CDM cosmology and the EDE cosmology are in tension with $S_8$ measurements, it is reasonable to ask whether their could exist additional extensions that would help in accommodating the low $S_8$ value. 
Naturally one can argue that `Occam's razor' should prevail, and that the `true' solution should be able to resolve both tension simultaneously. This might very well be the case, and it is without question that extending the parameter space until data fits is not a reasonable attitude. However, within a phenomenological framework, what is reasonable is to understand what aspects of a suggested resolution (EDE here) makes it at odds with a certain data set ($S_8$ here). This question is especially interesting given that the fiducial $\Lambda$CDM cosmology is equally at odds with these data. In other words, this approach does not consist in `hiding' bad effects of the EDE that compromises it with respect to $\Lambda$CDM in light of $S_8$ measurements. Rather, it consists in trying to understand what it would take (within a reasonable set of extensions) to accommodate current $S_8$ measurements in an EDE cosmology. If this can be achieved, the hope is then that it will lead to a set of predictions to be tested in the future, together with guidelines for model building which should make the extension less `ad-hoc'. 

 As was discussed extensively in this paper (see sec.~\ref{sec:sims}), it is interesting to note that at fixed $\omega_{\rm cdm}$, the EDE leads to a {\em decrease} in power at small scales that goes in the right direction to resolve the $S_8$ tension. However, the problem of EDE cosmologies is that they exploit a degeneracy with $\omega_{\rm cdm}$ to counteract the effect of the EDE on the gravitational potential wells as seen in the CMB.  Taken at face value, this poses both a concrete experimental problem --it is at odds with $S_8$ measurements-- and a theoretical `tuning' issue-- why should these two apparently unrelated sectors `conspire' to hide the EDE in CMB data?  This logic should also be applied when considering sensible extension to a model. 
One of the less `theoretically costly' possible explanation of the $S_8$ values is to invoke the fact that neutrinos are massive, and lead to a power suppression at small scales which decreases the value of $\sigma_8$. Unfortunately it seems as though in practice the required sum of neutrino masses $\sum m_\nu\sim0.3$eV is excluded by \Planck~data. Moreover, neither \DES~nor \KV~seems to have a preference for non-zero $\sum m_\nu$. 
However, it is interesting to note that constraints on the sum of neutrino masses can be strongly relaxed in extended cosmologies (e.g. \cite{Aghanim:2018eyx,Chacko:2019nej,Oldengott:2019lke}). In fact, in models attempting at resolving the Hubble tension with strongly interacting neutrinos \cite{Lancaster:2017ksf,Oldengott:2017fhy,Kreisch:2019yzn,Ghosh:2019tab}, it has been noted that Planck temperature 2015 data are in good agreement with $H_0\simeq 72$ km/s/Mpc and neutrino masses $\sum m_\nu\sim0.4$eV, but polarization data seems to restrict this resolution. In the EDE context, we have already mentioned that in Ref.~\cite{Sakstein:2019fmf}, it was suggested that the non-relativistic transition of neutrinos could trigger a phase-transition in the EDE. It would therefore be very interesting in the future to study further the possible connection between EDE and neutrino masses. Along this idea, we have performed a MCMC run in the 1-parameter EDE model against all data-sets with the sum of neutrino masses let free to vary. We find that marginalizing over the neutrino mass does not affect the result presented here. The $S_8$ value shifts downward only by $\sim0.3\sigma$. This is because, while $\sigma_8$ does get suppressed, a non-zero $\sum m_\nu$ increases $\Omega_m$, resulting only in a mild decrease in $S_8$. 

Another promising category of solutions to the $S_8$ tension invokes interaction between DM and an additional dark radiation (DR) component \cite{Lesgourgues:2015wza,Buen-Abad:2017gxg,Raveri:2017jto}. These models are particularly interesting because, by themselves and similarly to EDE, an additional radiation component leads to an increase in $S_8$. This is due to the fact that increasing the radiation density requires a simultaneous increase in the matter density to avoid a shift in matter radiation equality. As a consequence, $\Omega_m$ is higher in this model, leading to a higher $S_8$. However, the introduction of an interaction between DM and DR leads to a power-suppression at small-scales and therefore to a smaller $\sigma_8$, resulting in a net decrease in $S_8$. Alternatively, there exists a number of model leading to interactions between DM and DE at late-times. Introducing an interaction between DM and EDE is a straight-forward extension of the naive EDE model studied here. In fact, an axion EDE model whose dynamics is dictated by an interaction with dark gauge bosons was recently proposed in Ref.~\cite{Berghaus:2019cls}. Interestingly, the parameter space is not enlarged in this model- rather, the critical redshift $z_c$ at which the field starts to move is dictated by the ratio of the interaction rate over the Hubble rate. It will be interesting to generalize the model studied there (i.e., consider different type of interactions) and include linear cosmological perturbations, to check whether the presence of the additional interaction could open a new degeneracy direction (alternative to the $f_{\rm ede}-\omega_{\rm cdm}$ one), which would prevent an increase in $S_8$.
Speculating further, if the presence of EDE is confirmed in the future, it is likely that it is connected to the existence of DE today, and perhaps even inflation. As a matter of fact, EDE models were introduced over a decade ago to alleviate the cosmological coincidence problem -- the fact that the dark energy density  and the matter density are very close from one another just today \cite{Griest:2002cu,Kamionkowski:2014zda}. Therefore, it is quite natural to ask whether the current epoch of accelerated expansion (and inflation) could be due to a dynamical scalar-field similar to the EDE, if more eras of such type can have occurred at other moments in the history of the Universe, and what would be the impact on the cosmic structure growth. In fact, in Ref.~\cite{Joudaki:2016kym}, it was shown that a time-evolving equation of state for DE is favored over $\Lambda$CDM from a combination of \Planck, {\sc KiDS-450} and \SHOES~data. However, these simple solutions are severely constrained by \BAO~and \Pantheon~data. In future work, it will be interesting to check whether these constraints can be affected by the presence of EDE, and whether a more complete picture for early and late dark energy can help restoring cosmological concordance.

\section{Conclusions}
\label{sec:concl}
In this paper, we have reassessed the viability of the EDE against a host of high- and low-redshift measurements, by combining LSS observations from recent weak lensing surveys \KV~and \DES~with \Planck~2018 CMB data, {\sc BOSS-DR12} BAO and growth function measurements, and the \Pantheon~compilation of luminosity distance to SNIa. 
Our results can be summarized as follows:
\begin{enumerate}
    \item[\textbullet] Within a phenomenological 3-parameters EDE model, we confirm that \Planck+\BAO+\FS+\Pantheon+\SHOES~favor $f_{\rm EDE}(z_c)\simeq0.1\pm0.03$, $z_c
    \simeq 4000^{+1400}_{-500}$ and $\Theta_i = 2.6^{+0.4}_{-0.03}$, with a $\Delta\chi^2 = -18.7$  compared to $\Lambda$CDM fitted on the same data set (i.e. a $\sim3.6\sigma$ preference over $\Lambda$CDM\footnote{We assume Gaussian posteriors with 3 additional parameters for simplicity.}). The inclusion of the latest \Planck~data (and in particular the more precise polarization measurements) does not spoil the success of the EDE resolution to the Hubble tension. When compared to the `concordance' $\Lambda$CDM model (i.e.~obtained from analysis without \SHOES~data), the EDE cosmology fits  \Planck+\BAO+\FS+\Pantheon~equally well, but can additionally accommodate the high local $H_0$ values. 
    \item[\textbullet] Following the approach of Ref.~\cite{Niedermann:2020dwg}, we have then shown that {\em reducing} the parameter space to a 1-parameter EDE model by fixing \Logzc~and $\Theta_i$ to their best fit values as obtained from a \Planck~data only analysis -- which strikingly coincide with those from the combined analysis with \SHOES~-- leads to $\sim 2\sigma$ preference for non-zero EDE, namely $f_{\rm EDE}(z_c)\simeq0.08\pm0.04$ from \Planck~CMB data alone. In this cosmology, the inferred $H_0\simeq70\pm1.5$ km/s/Mpc is in agreement at better than 2$\sigma$ with its local measurement from SH0ES. The addition of \BAO, \FS~ and \Pantheon~data has no significant impact on the result. Including a prior on $H_0$ from \SHOES~pulls up the reconstructed fraction to the $\sim10\%$ level, with $H_0\simeq71.7\pm1$, while the fit to \Planck~is slightly better than in the concordance $\Lambda$CDM cosmology ($\Delta \chi^2 \sim -5$).
    \item[\textbullet] To justify the inclusion of LSS data in our analyses, we have confronted the EDE non-linear matter power spectrum as predicted by standard semi-analytical algorithms against a dedicated set of $N$-body simulations. We have then tested the 1-parameter EDE cosmology against WL data, finding that it does not significantly worsen the fit to the $S_8$ measurements as compared to $\Lambda$CDM, and that current WL observations do not exclude the EDE resolution to the Hubble tension. 
    \item[\textbullet] We also caution against the interpretation of constraints obtained from combining \Planck~with \KV+\DES. As we showed, the `compromise' cosmology that is obtained is a poor fit to \KV+\DES~and degrades the fit to \Planck~data, even in $\Lambda$CDM. This illustrates that these data sets are statistically inconsistent in a $\Lambda$CDM framework, and it is easily conceivable that the resolution of this tension lies elsewhere (whether systematic effect or new physics).
    \item[\textbullet] In light of the CMB lensing anomaly, we have shown that the lensing-marginalized CMB data favor non-zero EDE at $\sim2\sigma$, predicts $H_0$ in $1.4\sigma$ agreement with \SHOES~and $S_8$ in $1.5\sigma$ and $ 0.8\sigma$ agreement with \KV~and \DES,~respectively. There still exists however a $\sim2.5\sigma$ tension with the joint results from \KV~and~\DES. Moreover, the presence of EDE does not affect the amount of anomalous lensing. This suggests that the anomalous lensing is not due to the presence of EDE, but also that the success of EDE is not due to opening up a new degeneracy direction with some exotic lensing parameters. 
    \item[\textbullet]  With an eye on Occam's razor, we finally discussed extensions of the EDE cosmology that could allow to accommodate the low $S_8$ values. In particular, we argue that EDE models which are coupled to neutrinos; include interaction with an extra dark radiation bath or dark matter; or are connected to dynamical dark energy at late time (and perhaps inflation), are all worth exploring in future work as promising ways to fully restore cosmological concordance.  
\end{enumerate}

In another study~\cite{Smith:2020rxx}, we confronted BOSS data to the 1-param EDE cosmology within the EFT of LSS framework \cite{DAmico:2020ods,Ivanov:2020ril}, finding that the constraints on $f_{\rm EDE}$ largely weaken and that LSS current data do not exclude the EDE resolution to the Hubble tension.  
An important follow-up to these studies will be to see whether the new ACT data \cite{Aiola:2020azj}, compatible with \Planck~(although see Ref.~\cite{Handley:2020hdp}), support -- or restrict -- the EDE resolution to the Hubble tension. Looking forward, future CMB experiment (such as Simons Observatory \cite{Ade:2018sbj} and CMB-S4 \cite{Abazajian:2016yjj}) and LSS data (from Euclid \cite{Amendola:2016saw}, LSST \cite{Mandelbaum:2018ouv}, JWST and DESI \cite{Aghamousa:2016zmz}) will be crucial in testing prediction of the EDE cosmology (and its potential extensions) \cite{Smith:2019ihp,Klypin:2020tud} and firmly confirm -- or exclude -- the presence of EDE.

\section*{Acknowledgements}
The authors are extremely thankful to Tristan Smith for many thorough comments at various stages of this study, as well as Marc Kamionkowski, Kim Boddy, Jose L. Bernal, Tanvi Karwal, Rodrigo Calder\'on, Gabriele Parimbelli, Mikhail Ivanov and Colin Hill for interesting discussions. The authors thank Deanna C. Hooper, Julien Lesgourgues and Thejs Brinckmann for their help with the MontePython code. VP also thanks Nikita Blinov for his help with the implementation of the minimizer {\sc iMinuit} in MontePython.
The authors acknowledge the use of computational resources from the Ulysses SISSA/ICTP super-computer in Trieste, the CNRS/IN2P3 Computing Centre (CC-IN2P3) in Lyon, the IN2P3/CNRS and the Dark Energy computing Center funded by the OCEVU Labex (ANR-11-LABX-0060) and the Excellence Initiative of Aix-Marseille University - A*MIDEX, part of the French “Investissements d’Avenir” programme.

\appendix

\section{$\chi^2$ tables}
We report all $\chi^2_{\rm min}$'s obtained with the {\sc Minuit} algorithm \cite{James:1975dr} through the {\sc iMinuit} python package for the various model and data-set combination considered in this work.
\label{sec:Appendix_chi2}

\begin{table*}[t!]
\scalebox{0.9}{
  \begin{tabular}{|l|c|c|c|c|c|c|c|c|c|}
  \hline
   \multicolumn{10}{|c|}{$\Lambda$CDM cosmology} \\
  \hline
  \Planck~high$-\ell$ TT,TE,EE & 2347.86 & 2351.81 & 2347.02 & 2349.78 & $-$&2351.53 &2349.68 & 2351.24 & 2352.88  \\
\Planck~ low$-\ell$ EE &  396.03 &   395.8 & 399.60 & 395.71 & $-$&395.78 & 396.94& 395.88& 397.21 \\
\Planck~ low$-\ell$ TT &  23.18 & 22.25 & 22.74&  22.62&$-$ & 22.99 &22.83 & 22.34&22.09  \\
\Planck~lensing &$-$  & $-$ & 8.65&   9.56&$-$& 9.49 & 9.05& 9.93&10.05  \\
\Pantheon & $-$ & $-$& 1026.83&   1026.82&$-$& 1026.84 & 1026.69& 1026.72& 1026.67  \\
\BAO~\FS~BOSS DR12 &  $-$ & $-$&6.25& 6.11&$-$&  6.23 & 5.88 & 5.86 & 6.18 \\
\BAO~BOSS low$-z$ & $-$ & $-$&1.38& 1.39& $-$ & 1.34&  1.82 & 1.65 &  2.22\\
\SHOES &  $-$& 16.57 & $-$&18.57& $-$& $-$ &16.05&$-$ &  14.23\\
{\sc KiDS/Viking} &$-$ $-$& &$-$& $-$&177.9 &  182.62 & 182.21&$-$ & $-$  \\
{\sc COSEBI} & $-$ & $-$&$-$&$-$ & $-$& $-$ & $-$ & 8.30 & 6.44\\

  \hline
total & 2767.07&2786.43 & 3812.47& 3830.57 & 177.9 &  3996.82&4011.16& 3821.93&3837.98  \\
  \hline

  \end{tabular}  
  }
  \caption{Best-fit $\chi^2$ per experiment (and total) in the $\Lambda$CDM model.}
  \label{tab:chi2_lcdm}
\end{table*}
\begin{table*}[t!]
\scalebox{0.9}{
  \begin{tabular}{|l|c|c|c|c|}
  \hline
  \multicolumn{5}{|c|}{3-parameter EDE cosmology} \\
  \hline
  \Planck~high$-\ell$ TT,TE,EE & 2343.07 & 2350.24 & 2349.30&2347.73  \\
\Planck~ low$-\ell$ EE  &  397.47& 396.20& 398.19& 395.88 \\
\Planck~ low$-\ell$ TT  &21.54& 20.80 &20.56 &21.09  \\
\Planck~lensing  &$-$ & $-$&10.12 & 9.85\\
\Pantheon  &$-$ & $-$&1026.72 & 1026.68\\
\BAO~BOSS DR12  & $-$ &$-$ & 3.46&  $-$\\
\BAO~BOSS low$-z$  & $-$& $-$& 2.06& 1.81\\
\BAO/\FS~BOSS DR12  & $-$& $-$& $-$& 6.73\\
\SHOES  &$-$ & 0.47& 1.38&2.13 \\
  \hline
total &2762.08 &2767.72 &2786.43 & 3811.89 \\
  \hline

  \end{tabular} 
  }
  \caption{Best-fit $\chi^2$ per experiment (and total) in the 3-parameter EDE model.}
    \label{tab:chi2_ede}

\end{table*}

\begin{table*}[t!]
\scalebox{0.9}{
  \begin{tabular}{|l|c|c|c|c|c|c|c|c|c|}
  \hline
  \multicolumn{10}{|c|}{1-parameter EDE cosmology} \\
  \hline
  \Planck~high$-\ell$ TT,TE,EE & 2345.02 &2347.63 & 2344.98& 2347.42& $-$&2345.16 & 2349.15& 2350.22 & 2349.82 \\
\Planck~ low$-\ell$ EE  &  395.80& 395.97&395.82 &395.90 & $-$&396.33 & 395.88& 396.10 &395.79 \\
\Planck~ low$-\ell$ TT  &21.49& 20.82& 21.89& 20.85& $-$& 22.38&20.97 & 21.54 &  20.84\\
\Planck~lensing  & $-$& $-$&9.39& 10.00& $-$& 9.07& 10.04& 10.22&  10.91 \\
\Pantheon  & $-$&$-$& 1026.80 &1026.69&$-$ & 1026.84& 1026.7& 1026.69&1026.80\\
\BAO~\FS~BOSS DR12  & $-$&$-$&6.44 & 7.18&$-$ &6.43 & 7.08& 6.47 &7.70 \\
\BAO~BOSS low$-z$  & $-$&$-$& 1.41& 2.33&$-$ &1.33 & 2.33&2.38& 2.83\\
\SHOES  &$-$ & 1.07&$-$ & 1.64& $-$& $-$& 1.62&  $-$&  2.43\\
{\sc KiDS/Viking} &$-$ &$-$&$-$ &$-$  &178.0 &184.57  & 183.88& $-$& $-$  \\
{\sc COSEBI}  & $-$&$-$&$-$ &$-$ & $-$& $-$ & $-$ & 6.83 & 9.22\\

  \hline
total &2762.31 &2765.49 & 3806.74& 3812.01 & 178.0 &3992.11 & 3997.67&3820.46 & 3826.35 \\
  \hline

  \end{tabular} 
  }
  \caption{Best-fit $\chi^2$ per experiment (and total) in the 1-parameter EDE model.}
    \label{tab:chi2_1pede}

\end{table*}

\begin{table*}[!t]
\scalebox{0.9}{
  \begin{tabular}{|l|c|c|c|c|}
  \hline
  Model &  \multicolumn{2}{c|}{$\Lambda$CDM} & \multicolumn{2}{c|}{EDE cosmology} \\
  \hline
  \Planck~high$-\ell$ TT,TE,EE &2339.92 & 2340.96 & 2335.71&  2336.12\\
\Planck~ low$-\ell$ EE  & 395.67 & 395.87 & 395.80 & 397.01  \\
\Planck~ low$-\ell$ TT  & 21.93 & 21.03 & 20.65& 20.44\\
\Planck~lensing  &  8.47& 8.35 & 9.82& 9.36 \\
\Pantheon  & $-$&1026.88 &$-$  & 1026.99\\
\BAO~\FS~BOSS DR12  & $-$  & 8.04 &$-$  & 9.02 \\
\BAO~BOSS low$-z$  &  $-$& 3.20 &$-$  & 3.48\\
\SHOES  &   $-$&10.67 & $-$ &1.91 \\
{\sc COSEBI}  & $-$  & 1.2 &$-$  & 4.07 \\
  \hline
total &  2765.99& 3816.23& 2761.98 & 3808.40 \\
  \hline

  \end{tabular} 
  }
  \caption{Best-fit $\chi^2$ per experiment (and total) in $\Lambda$CDM and the 1-parameter EDE model when marginalizing over the lensing information in Planck.}
    \label{tab:chi2_edelcdm}

\end{table*}

\section{a closer look to $N$-Body simulations}\label{ap:sims}

\begin{figure*}
\centering
\begin{tabular}{cc}
\includegraphics[width=.499\textwidth]{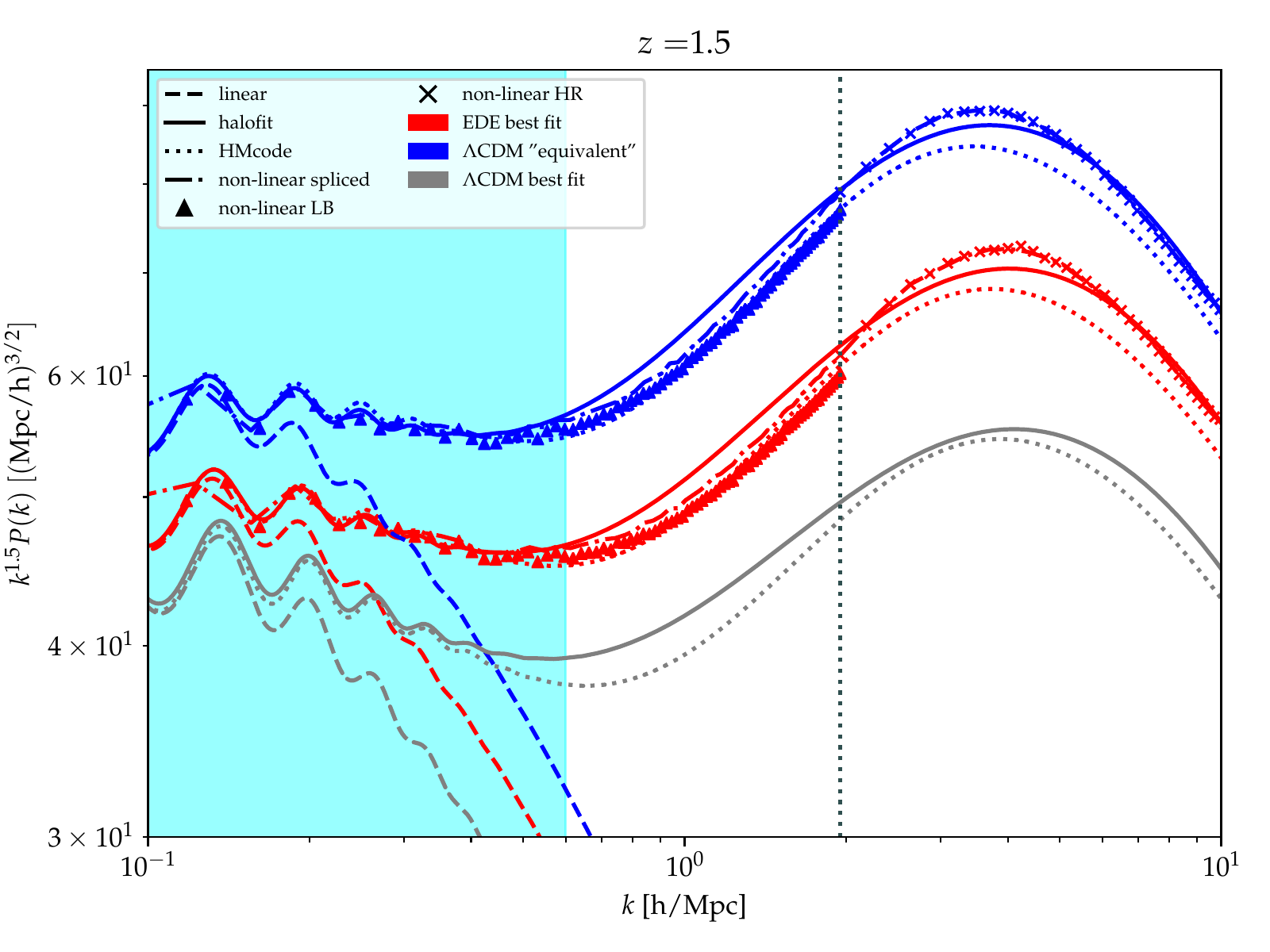} &
\includegraphics[width=.499\textwidth]{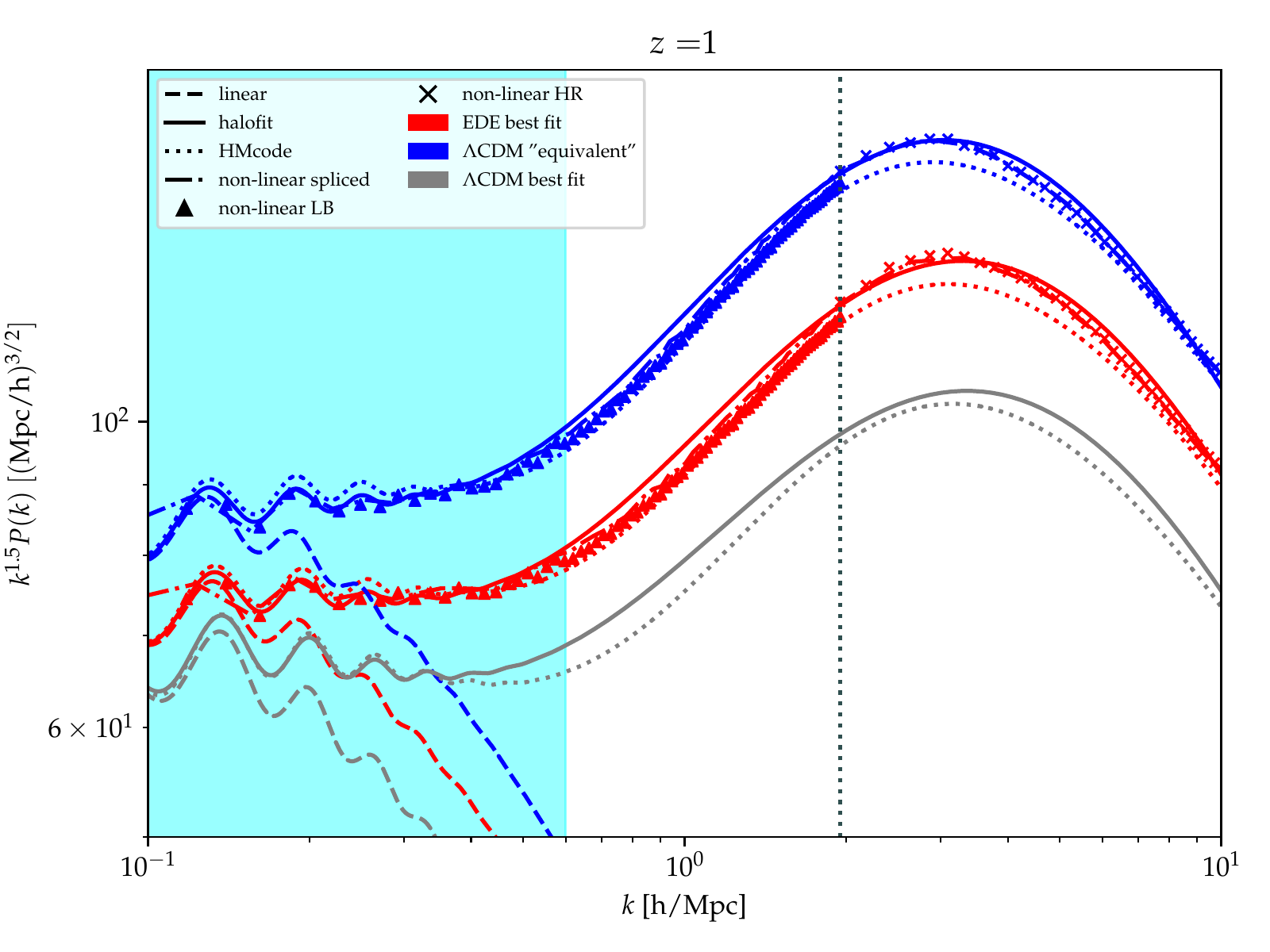} \\
\end{tabular}
\includegraphics[width=.499\textwidth]{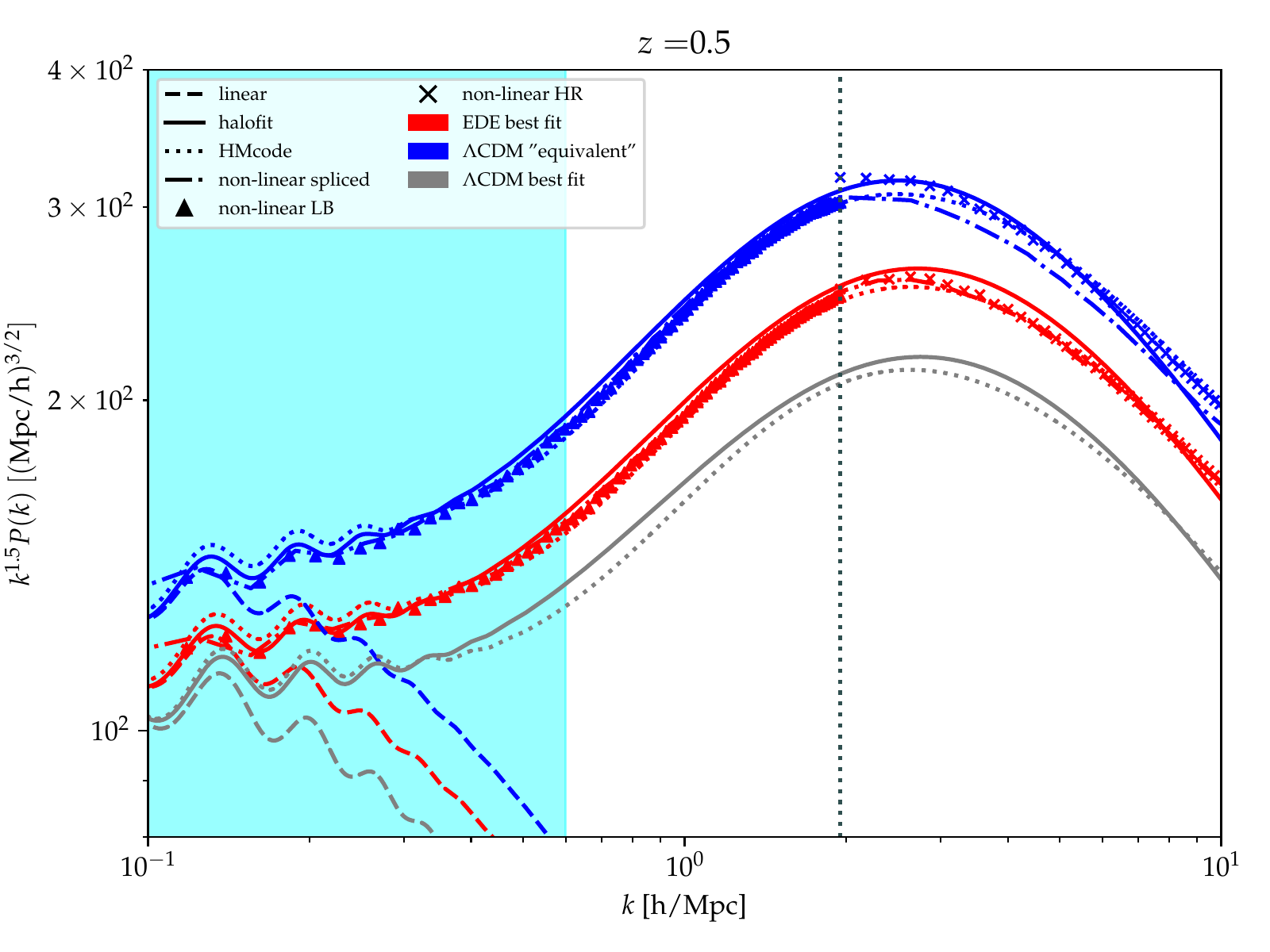}
\caption{\looseness=-1 Here we compare the matter power spectra
extracted from our simulations, with the ones computed with {\sc{halofit}}/{\sc{HMcode}}, in three different redshift bins from $z=1.5$ to $z=0.5$.
The blue curves refer to the $\Lambda \rm CDM$ scenario, whereas the red ones refer to the EDE best fit model. As a reference, we also report the best fit $\Lambda \rm CDM$ case from Planck 2018.
The spliced power spectra are denoted by thick dot-dashed lines. Symbols stand for the output power spectra of the ``non-spliced'' LB and HR simulations.
The solid/dotted lines are the non-linear power spectra from {\sc{halofit}}/{\sc{HMcode}}, while the dashed lines are the corresponding linear power spectra used to set the initial conditions for the simulations. The cyan shaded band approximately corresponds to the scales probed by DES-Y1.}
\label{fig:pk_abs}
\end{figure*}

\begin{figure*}
\centering
\begin{tabular}{cc}
\includegraphics[width=.499\textwidth]{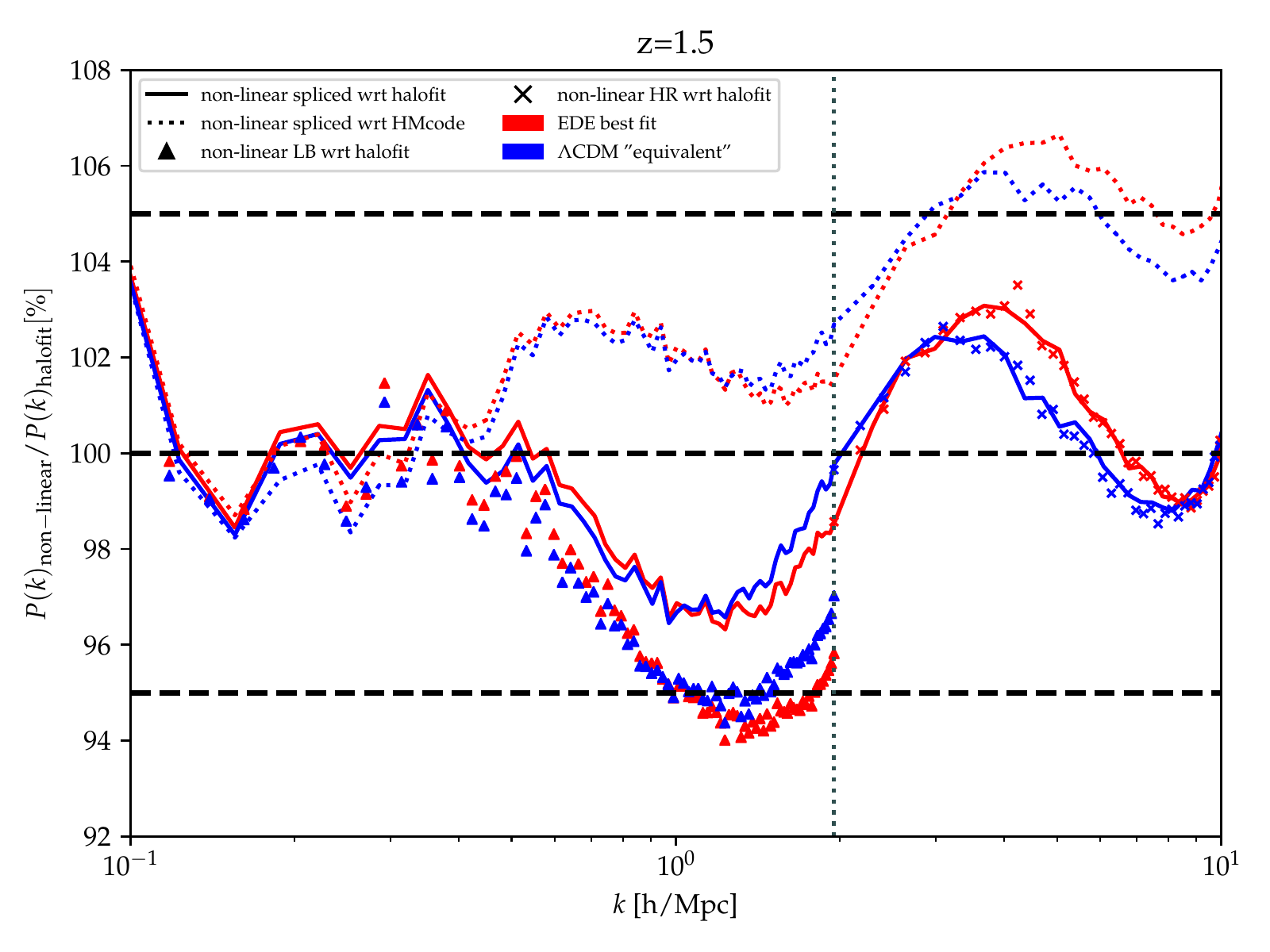} &
\includegraphics[width=.499\textwidth]{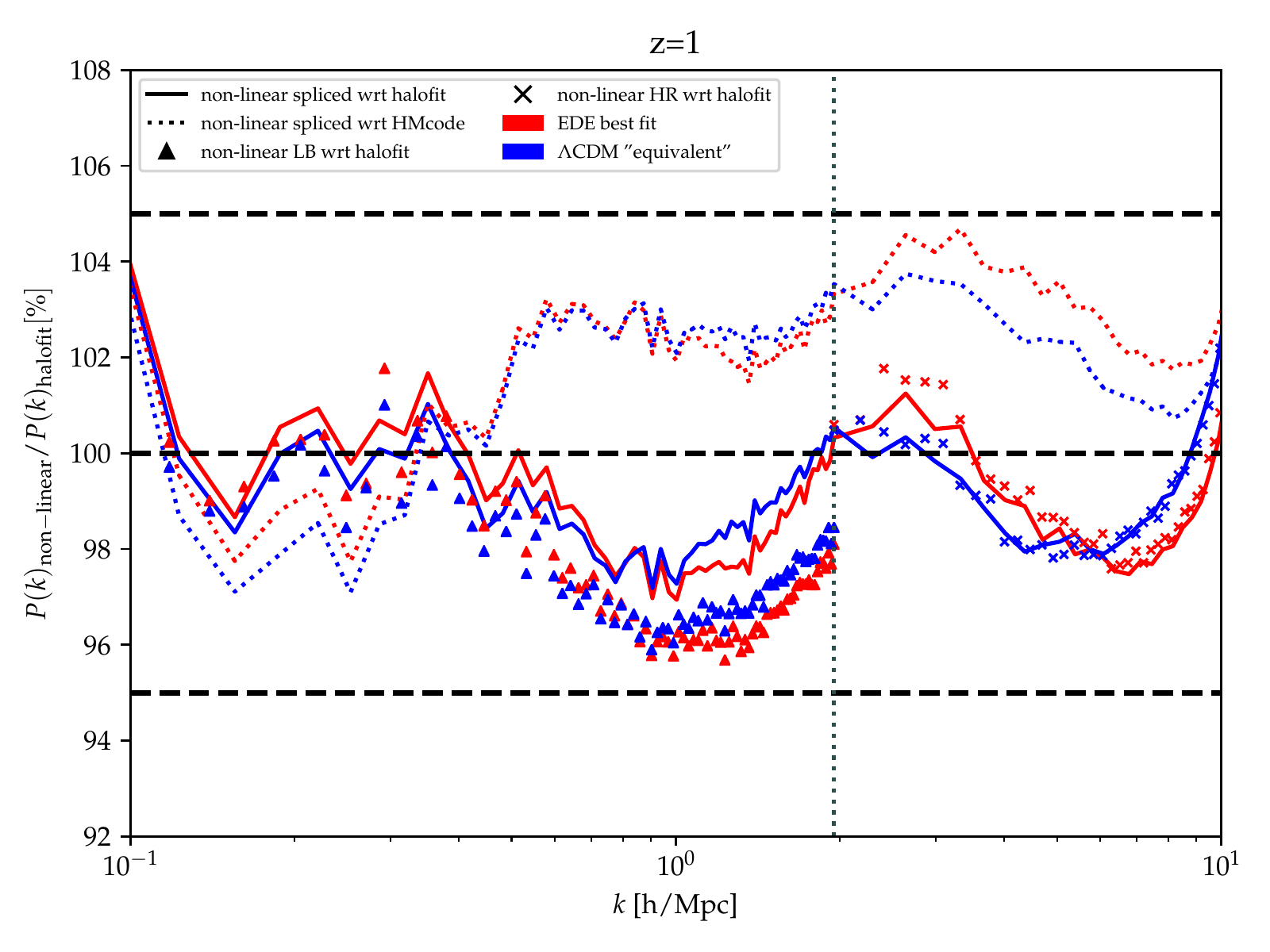} \\
\end{tabular}
\includegraphics[width=.499\textwidth]{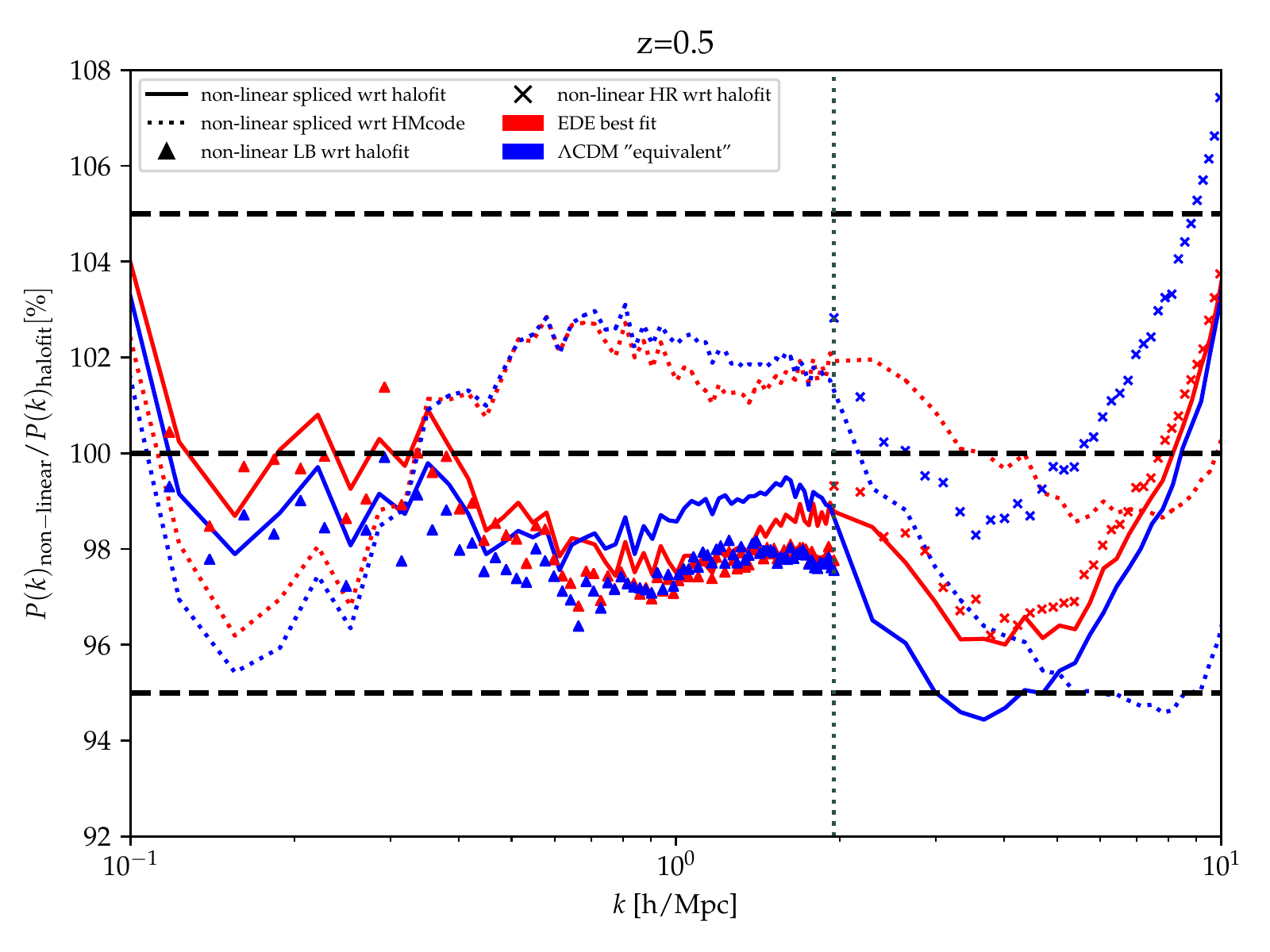}
\caption{\looseness=-1 Here we show the ratio between the non-linear matter power spectra from our simulations and the ones computed with {\sc{halofit}}/{\sc{HMcode}}, for both the \LCDM~equivalent and the EDE best-fit models. We have adopted the same linestyle-code and color-code of Figure \ref{fig:pk_abs}.}
\label{fig:ratio_hf}
\end{figure*}

\begin{figure*}
\centering
\begin{tabular}{cc}
\includegraphics[width=.499\textwidth]{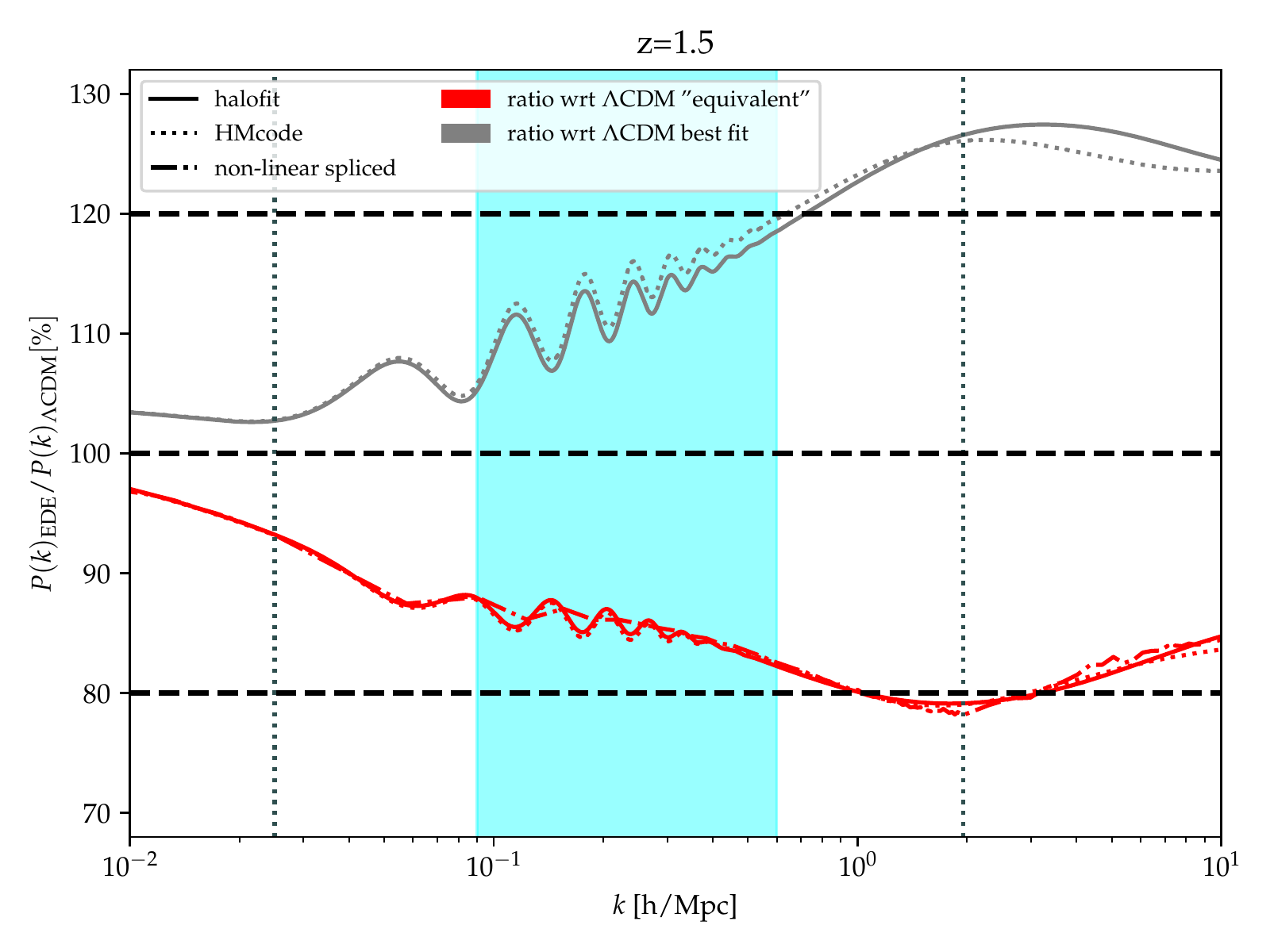} &
\includegraphics[width=.499\textwidth]{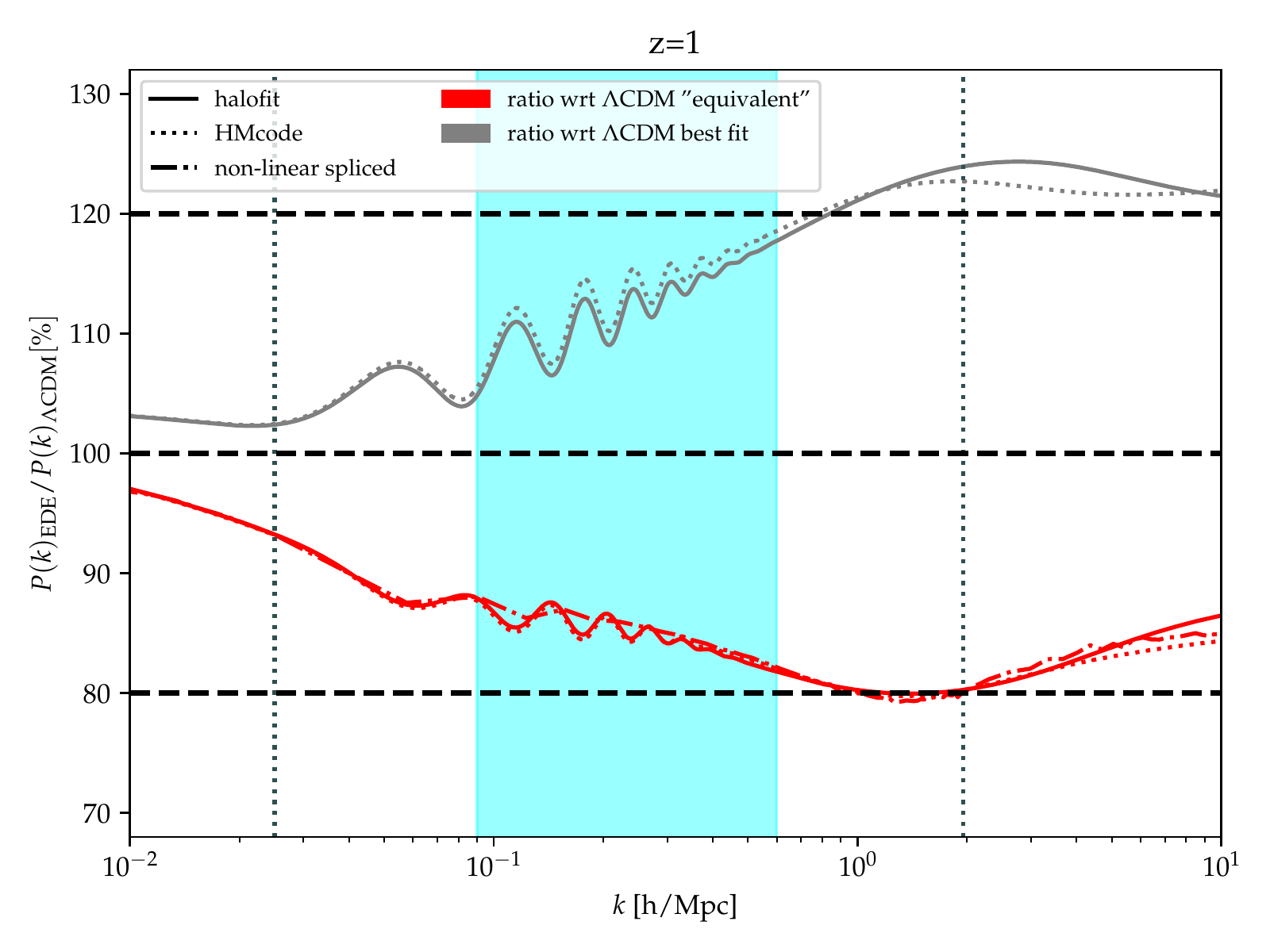} \\
\end{tabular}
\includegraphics[width=.499\textwidth]{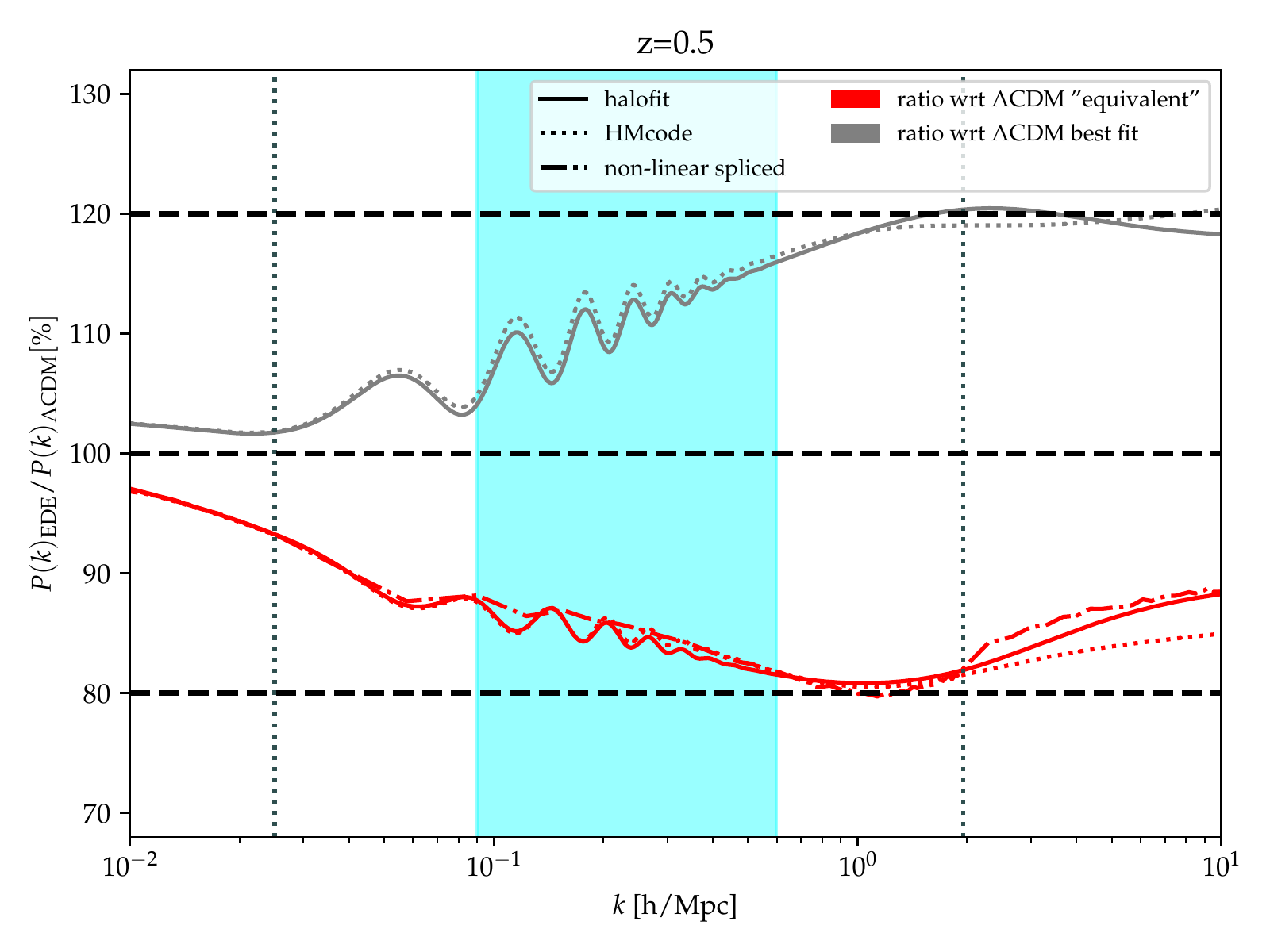}
\caption{\looseness=-1 Here we compare departures from the \LCDM~``equivalent'' model in terms of ratios of non-linear matter power spectra. The EDE best-fit case is shown in red, while the gray lines refer to the \LCDM~best-fit model.
Solid and dotted lines stand for the non-linear power spectra from {\sc{halofit}} and {\sc{HMcode}}, respectively.
Dot-dashed lines refer to the outputs of our simulations. The cyan shaded band approximately corresponds to the scales probed by DES-Y1.}
\label{fig:ratio_lcdm}
\end{figure*}

\begin{figure*}
\centering
\begin{tabular}{cc}
\includegraphics[width=.499\textwidth]{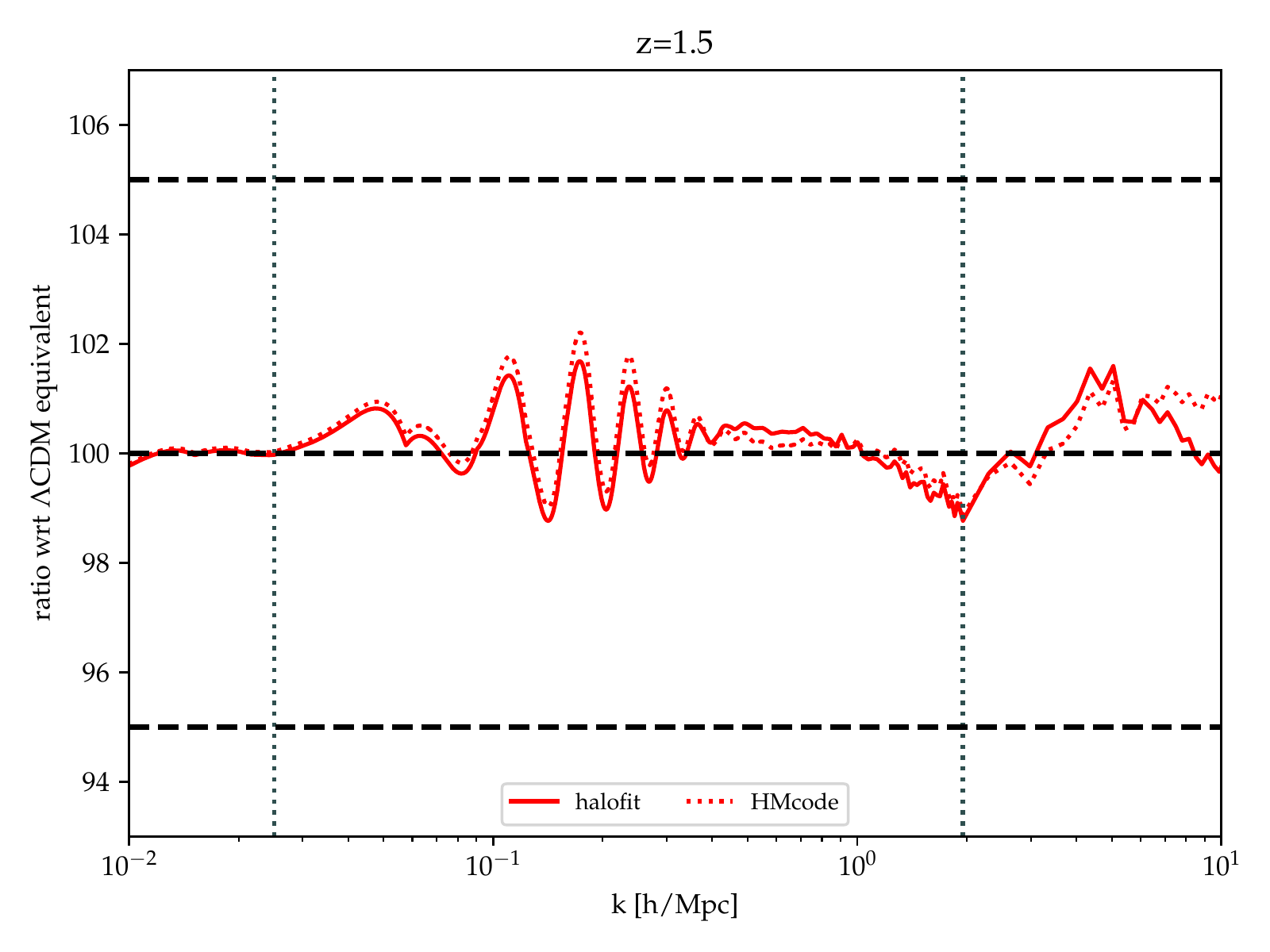} &
\includegraphics[width=.499\textwidth]{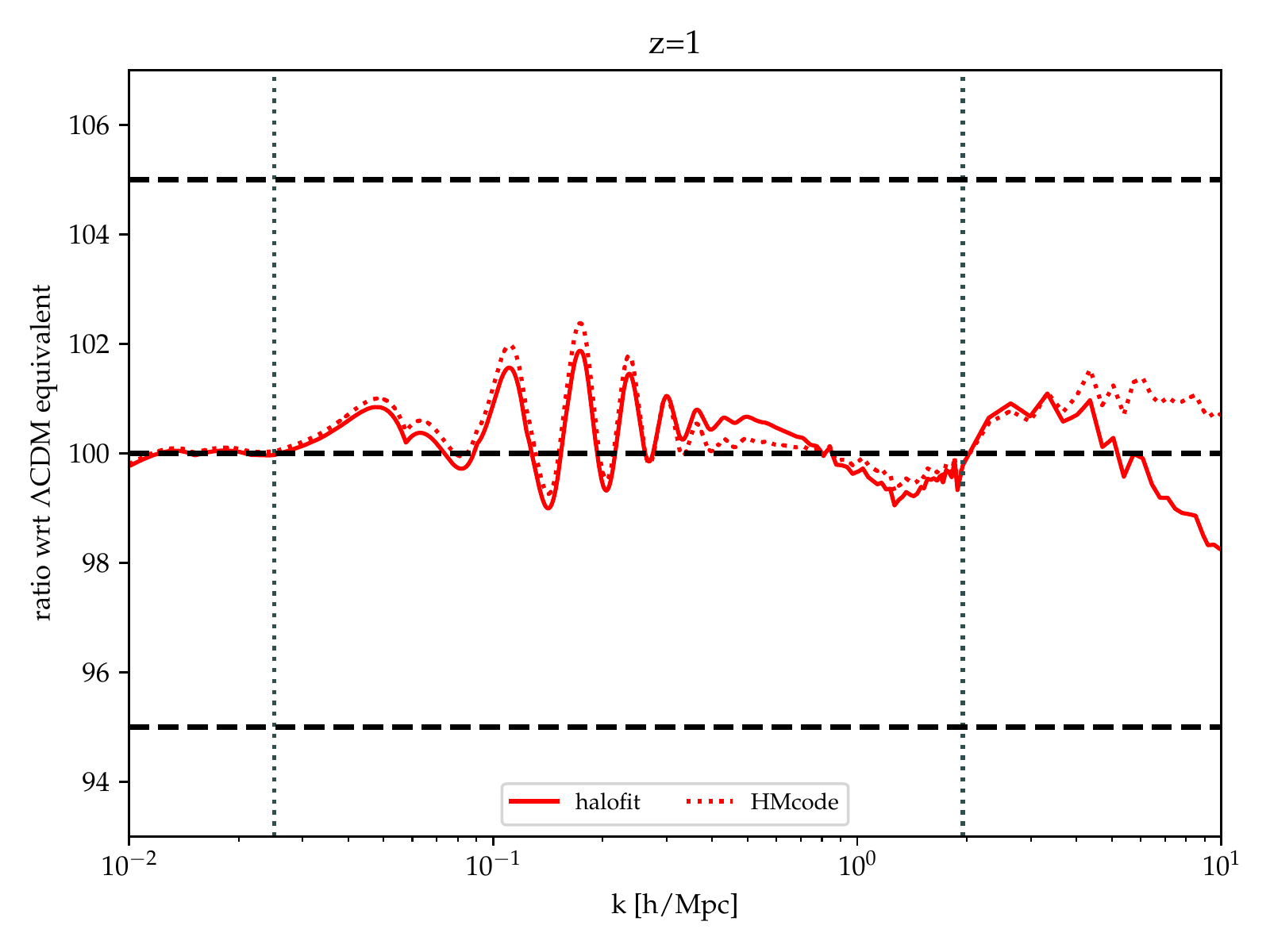} \\
\end{tabular}
\begin{tabular}{cc}
\includegraphics[width=.499\textwidth]{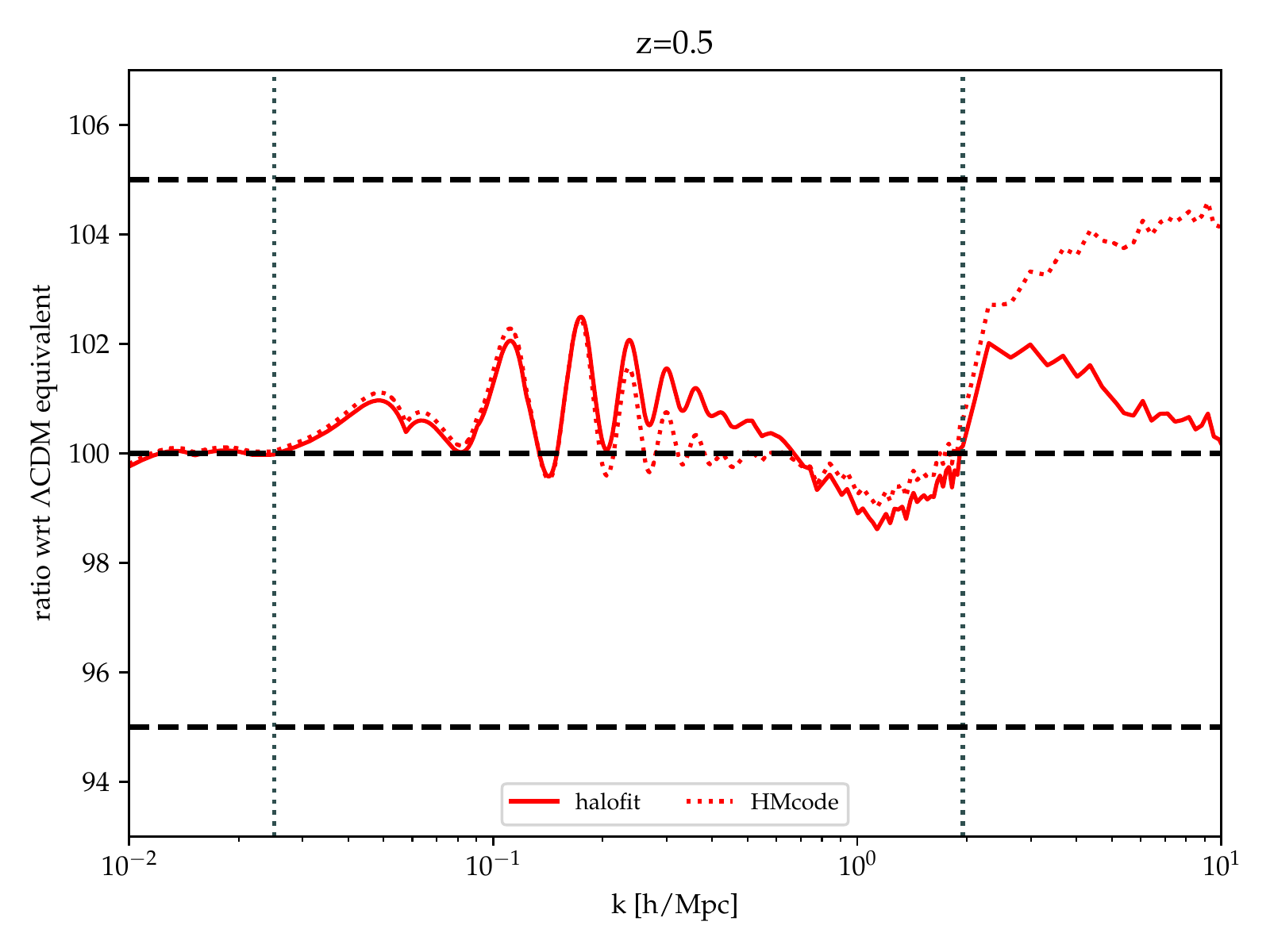}
\includegraphics[width=.499\textwidth]{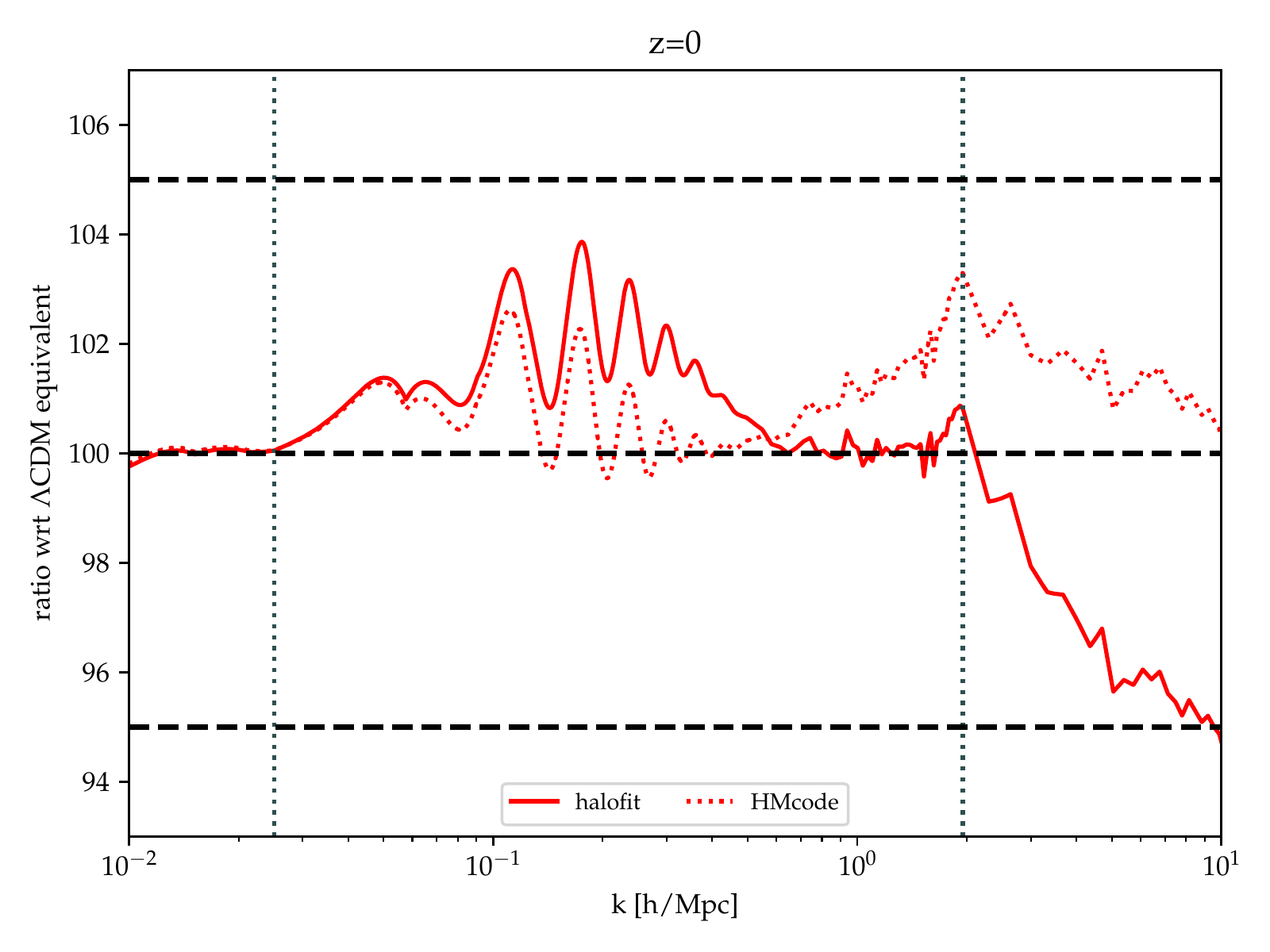}
\end{tabular}
\caption{\looseness=-1 Here we compare the outputs of our simulations with the {\sc{halofit}}/{\sc{HMcode}} predictions, in terms of deviations in the ratios of the EDE best-fit power spectra over the \LCDM~``equivalent'' ones, in four redshift bins from $z=2$ to $z=0$.
Solid and dotted lines stand for {\sc{halofit}} and {\sc{HMcode}}, respectively.} 
\label{fig:ratio_lcdm_hf}
\end{figure*}

The main systematical uncertanties in numerical simulations come from their limited box size and resolution, as it has been thoroughly discussed in past literature (see, e.g., \cite{Heitmann:2008eq,Rasera:2013xfa,Schneider:2015yka,Klypin:2018ydv}). In order to minimize the missing large-scale modes, potentially affecting small-box simulations, and to overcome the impossibility of capturing the very non-linear scales in our large-box simulations, we adopted a splicing technique to bind together the matter power spectra extracted from simulations with different resolutions, for each redshift and model, as in Refs.~\cite{McDonald:2001fe,Borde:2014xsa}. 

For both the $\Lambda$CDM model and the EDE best fit model from \cite{Smith:2019ihp}, we indeed performed: i) one Large Box (LB hereafter) simulation with $N = 1024^3$ DM particles and box size $L = 250 \, h^{-1} {\rm Mpc}$; 
ii) one High Resolution (HR hereafter) simulation with $N = 1024^3$ DM particles and box size $L = 1000 \, h^{-1} {\rm Mpc}$;
iii) one Low Resolution (LR hereafter) simulation with the same box size of the HR one and the same resolution of the LB, namely $N = 256^3$ and $L = 250 \, h^{-1} {\rm Mpc}$, to be used as a transition simulation.
The spliced non-linear matter power spectrum $P(k)$ is given by Ref.~\cite{Borde:2014xsa}
\begin{equation}
\displaystyle
P(k)=
    \begin{cases}
      P_{\rm LB}(k) \cdot \frac
      {\displaystyle P_{\rm HR}(k_{\rm MIN}^{\rm 250})}
      {\displaystyle P_{\rm LR}(k_{\rm MIN}^{\rm 250})},
      & \text{if}\ k \leq k_{\rm MIN}^{\rm 250} \\ 
      P_{\rm LB}(k) \cdot \frac
      {\displaystyle P_{\rm HR}(k)}
      {\displaystyle P_{\rm LR}(k)}, 
      & \text{if}\ k_{\rm MIN}^{\rm 250} < k < \frac{1}{2}k_{\rm Nyq}^{\rm LB} \\ 
      P_{\rm HR}(k) \cdot \frac
      {\displaystyle P_{\rm LB}(0.5 \cdot k_{\rm Nyq}^{\rm LB})}
      {\displaystyle P_{\rm LR}(0.5 \cdot k_{\rm Nyq}^{\rm LB})},
      & \text{if}\ k \geq 0.5 \cdot k_{\rm Nyq}^{\rm LB}
    \end{cases}
\label{eq:splicing}
\end{equation}

where $k_{\rm MIN}^{\rm 250}$ is the minimum $k$-mode in our small-box simulations (HR and LR), while $k_{\rm Nyq}^{\rm LB}$ is the Nyquist wave-number of the LB one.

Besides the aforementioned systematical uncertanties, numerical simulations are also affected by two primary sources of statistical errors: the cosmic variance, affecting the large-scale part of the spectra, and the shot noise due to the discreteness of the DM particles, thereby affecting the smallest scales. 

Concerning the shot noise term, its contribution to the power spectrum is simply given by $P_{\rm SN} = ( L/N )^3$. 
It is straightforward to see that it is largely subdominant at the scales and redshifts considered in this work, from Fig.~\ref{fig:pk_abs}, where we compare the matter power spectra
extracted from our simulations with the ones computed with {\sc{halofit}}/{\sc{HMcode}}, in three different redshift bins from $z=1.5$ to $z=0.5$ -- given that we have already discussed the $z=0$ case in Section~\ref{sec:sims}. In Fig.~\ref{fig:ratio_hf} we plot the ratio between the power spectrum predicted by  {\sc{halofit}} or {\sc{HMcode}} and that extracted from the numerical simulation in order to explicitly demonstrate that the differences are below $5\%$ level, for scales $10^{-2} \lesssim  k \lesssim 10$~$h/$Mpc, at redshifts $ 0.5 \leq z \leq 2$, for both $\Lambda \rm CDM$ and EDE models. This extends the $z=0$ result presented in the main text to cover the full redshift range from \KV.

It is also informative to compare the prediction from algorithms with $N$-body at larger scales than that depicted in  Figs.~\ref{fig:pk_abs} and~\ref{fig:ratio_hf}. Indeed, these are affected by higher statistical noise, due to cosmic variance, as one might already guess from the lower$-k$ part of both figures. To beat down cosmic variance, one should run several statistical realizations of the same simulation, by producing initial conditions starting from different random seeds.
To circumvent this issue and save computational time, we adopted the simple solution to run the two sets of simulations (EDE and $\Lambda$CDM) with identical random seeds for the realization of their initial conditions, and to present our results in terms of ratios in the matter power spectra between the EDE and the $\Lambda \rm CDM$ models, in Figs.~\ref{fig:ratio_lcdm} and~\ref{fig:ratio_lcdm_hf}. Any scatter related to the cosmic variance is now removed, allowing us to go down one order of magnitude in terms of wave-numbers $k$'s. We show in both figures by vertical dashed lines the scales corresponding to $k_{\rm MIN}^{\rm 250}$ and $k_{\rm Nyq}^{\rm LB}/2$. One can clearly see again that the EDE by itself lead to a decrease in power. However, the increase in $\omega_{\rm cdm}$, leads the EDE best-fit model to predict $~\mathcal{O}(20\%)$ increase in power, when compared to the best-fit $\Lambda$CDM model. Note how the differences become even more manifest at higher redshift. This illustrates that high-$z$ LSS measurements have the potential to put EDE under crucial tests \cite{Klypin:2020tud}.

Another way of presenting our results is in terms of the accuracy at which {\sc{halofit}}/{\sc{HMcode}} can predict {\em deviations} in the non-linear power spectrum of EDE models with respect to the $\Lambda \rm CDM$ `equivalent' case (as opposed to predicting the absolute power spectrum). This is what we show in Fig.~\ref{fig:ratio_lcdm_hf}, where we now compare the ratio between the EDE and $\Lambda \rm CDM$ power spectra from {\sc{halofit}}/{\sc{HMcode}} against the same ratio extracted from simulations. The thick horizontal lines highlight $\pm 5\%$ deviations. In light of all of this, we conclude that, in the EDE framework, {\sc{halofit}}/{\sc{HMcode}} predictions on $\Lambda \rm CDM$ departures are reliable at $\leq 5\%$ level with respect to the outputs of $N$-Body simulations, for scales $10^{-2} \lesssim  k \lesssim 10 ~h/$Mpc, at redshifts $ 0 \leq z \leq 2$.


\bibliography{bibliography}
\end{document}